\documentclass[prd,aps,floatfix,preprint,nofootinbib]{revtex4}
\usepackage{grffile}
\usepackage{graphicx}
\usepackage{amssymb}
\usepackage{amsmath}
\usepackage{hyperref}
\usepackage[dvipsnames]{xcolor}

\def\simge{
    \mathrel{\rlap{\raise 0.511ex
        \hbox{$>$}}{\lower 0.511ex \hbox{$\sim$}}}}
\def\simle{
    \mathrel{\rlap{\raise 0.511ex 
        \hbox{$<$}}{\lower 0.511ex \hbox{$\sim$}}}}

\begin{document}

\begin{abstract}
{We report on the computation of the connected light-quark vacuum polarization with 2+1+1 flavors of highly improved staggered quarks [Follana \textit{et al.}, \href{https://doi.org/10.1103/PhysRevD.75.054502}{Phys. Rev. D \textbf{75}, 054502 (2007).}] fermions at the physical point and its contribution to the muon anomalous magnetic moment. Three ensembles, generated by the MILC collaboration, are used to take the continuum limit. The finite-volume correction to this result is computed in the (Euclidean) time-momentum representation to next-to-next-to-leading order (NNLO) in chiral perturbation theory. We find \mbox{$a_\mu^{ll}({\rm HVP})=(659\pm 20\pm 5\pm 5\pm 4)\times 10^{-10}$}, where the errors are statistical and estimates of residual uncertainties from taking the continuum limit, scale setting, and truncation of chiral perturbation theory at NNLO. We compare our results with recent ones in the literature.}
\end{abstract}

\title{Light quark vacuum polarization at the physical point and contribution to the muon \begin{boldmath}$g-2$\end{boldmath}}

\author{Christopher Aubin}
\affiliation{Department of Physics and Engineering Physics, Fordham University, Bronx, New York, New York 10458, USA}
\author{Thomas Blum}
\author{Cheng Tu}
\affiliation{Physics Department, University of Connecticut, Storrs, Connecticut 06269-3046, USA}
\author{Maarten Golterman}
\affiliation{Department of Physics and Astronomy, San Francisco State University, San Francisco, California 94132, USA}
\author{Chulwoo Jung}
\affiliation{Physics Department, Brookhaven National Laboratory, Upton, New York 11973, USA}
\author{Santiago Peris}
\affiliation{Department of Physics and IFAE-BIST, Universitat Aut\`onoma de Barcelona, Barcelona, E-08193 Bellaterra, Spain}

\maketitle

\section{Introduction}

Fermilab experiment E989 is measuring the anomalous magnetic moment of the muon ($a_\mu=(g-2)/2$) with the goal of reducing the error on the BNL E821~\cite{Bennett:2006fi} result by a factor of 4. An upcoming experiment at J-PARC, E34, aims to do the same with a completely different technique. Lattice calculations of the hadronic contributions to the muon $g-2$, like the one reported here, are crucial to obtain and cross-check the standard model value to the same accuracy in order to discover new physics or lay to rest the longstanding discrepancy between theory and experiment.

In this paper we focus on the leading hadronic vacuum polarization (HVP) contribution to the muon anomaly. The aim is to test the efficacy of modern noise-reduction techniques to reduce the statistical errors of Monte Carlo methods used in lattice QCD in the context of the HVP and to provide accurate finite-volume corrections to these results using chiral perturbation theory at two-loop order.

\begin{figure}[htbp]
\begin{center}
\includegraphics[width=0.4\textwidth]{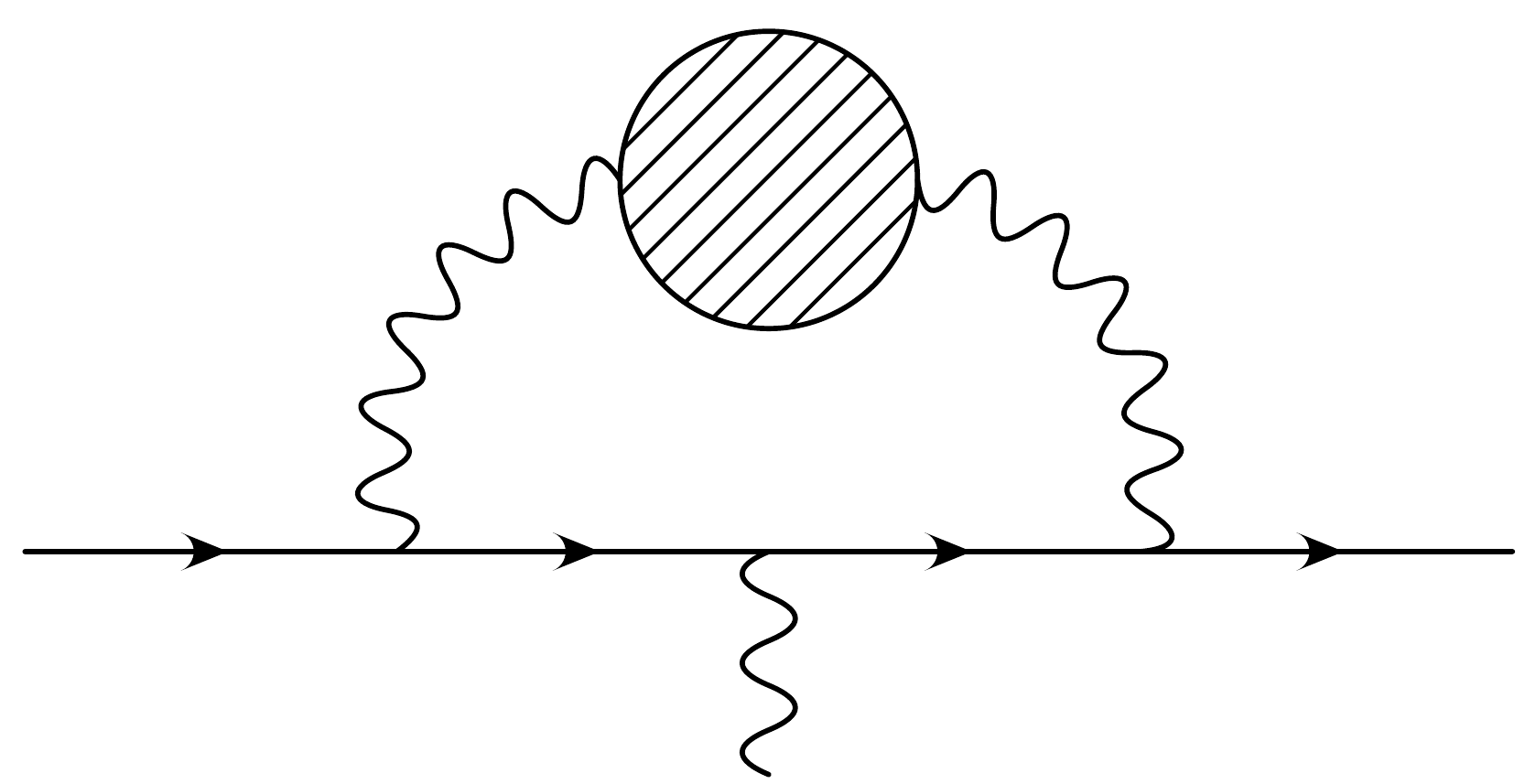}
\includegraphics[width=0.4\textwidth]{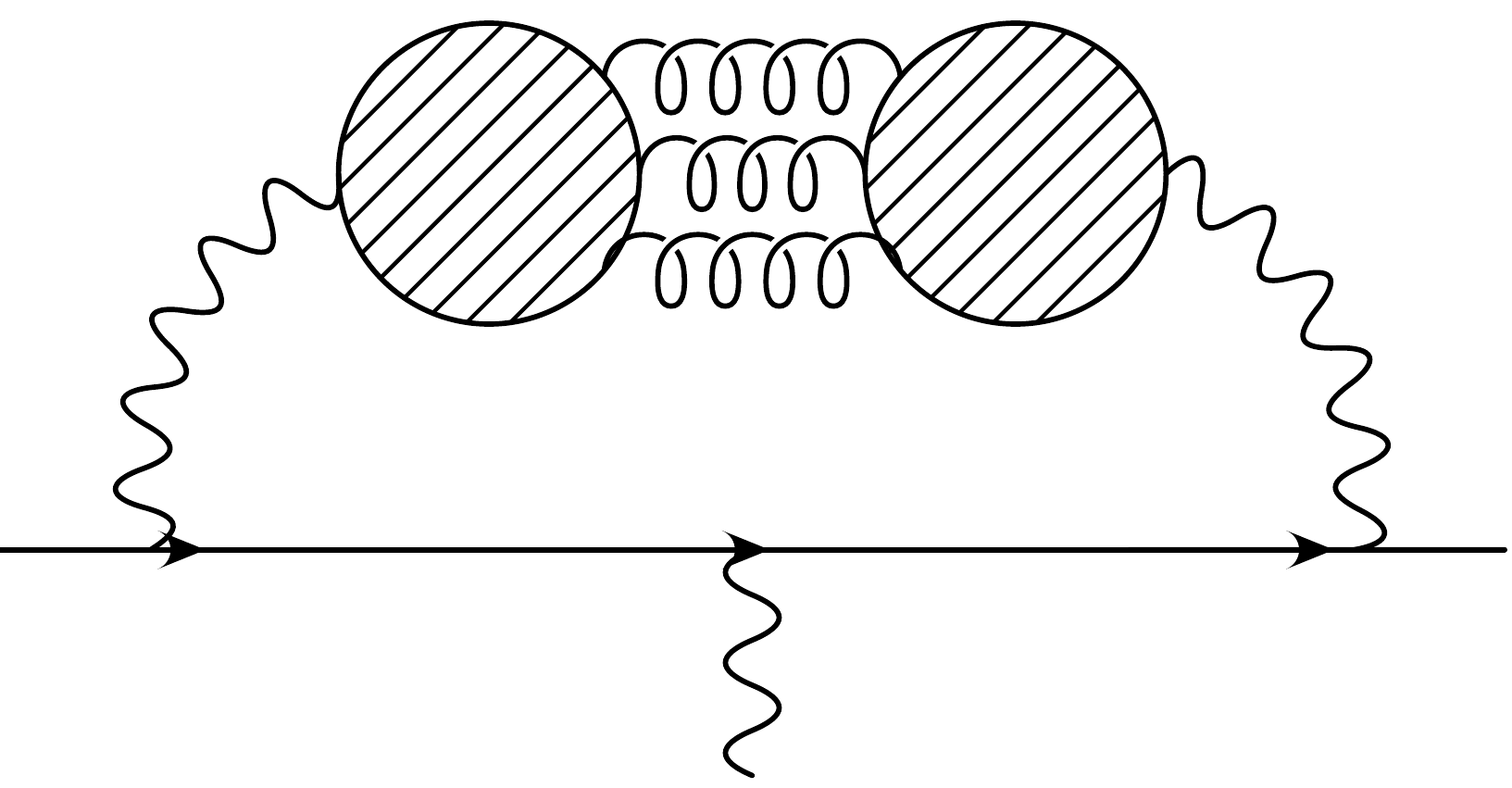}
\caption{The quark connected (left) and disconnected (right) diagrams contributing to the hadronic vacuum polarization contribution to the muon anomaly.}
\label{fig:vacpol}
\end{center}
\end{figure}

The total HVP contribution to $a_\mu$ comes from both connected- and disconnected-quark line diagrams shown in Fig.~\ref{fig:vacpol}, for each flavor of quark in Nature. The u, d quark-connected contributions are by far the largest, and we only compute them in this work. Comparison to other recent precise calculations~\cite{Borsanyi:2017zdw,Blum:2018mom,Giusti:2018mdh,Davies:2019efs,Shintani:2019wai,Gerardin:2019rua} will provide important validation for the lattice method.

The plan of this paper is the following. In Sec.~\ref{sec:th} we review the theoretical framework for the calculation, including important details of the lattice calculation and the calculation in chiral perturbation theory (ChPT) in Euclidean space of the leading and next-to-leading finite-volume corrections to the HVP contribution to the muon $g-2$. Section~\ref{sec:results} presents our results and comparison to other calculations. In Sec.~\ref{sec:conclusion} we give a summary of this work and discuss implications for future work and the important upcoming comparison with experiment. The appendix reports details of the next-to-next-to-leading order (NNLO) chiral perturbation theory calculation.

\section{Theoretical framework}
\label{sec:th}

Using lattice QCD and continuum, infinite-volume (perturbative) QED, one can calculate the hadronic vacuum polarization (HVP) contribution to the muon anomalous magnetic moment~\cite{Lautrup:1971jf,deRafael:1993za,Blum:2002ii},
\begin{eqnarray}
\label{eq:amu}
a_\mu^{\rm HVP} &=& 4\alpha^{2}\int_{0}^{\infty}d q^2 \, f(q^2)\,{\hat\Pi}(q^2),\\
f(q^2) &=& \frac{m_\mu^2 q^2 Z^3(1-q^2 Z)}{1+m_\mu^2 q^2 Z^2},\\
Z &=& -\frac{q^2-\sqrt{q^4+4 m_\mu^2 q^2}}{2 m_\mu^2 q^2}.
\end{eqnarray}
$m_\mu$ is the muon mass, and $\hat\Pi(q^{2})$ is the subtracted HVP, $\hat\Pi(q^{2})=\Pi(q^{2})-\Pi(0)$, computed directly on a Euclidean space-time lattice from the Fourier transform of the vector current two-point function,
\begin{eqnarray}
\label{eq:ft}
\Pi^{\mu\nu}(q) &=& \int d^4x\, e^{i q x}\langle j^\mu(x)j^\nu(0)\rangle\\
\label{eq:wi}
&=&\Pi(q^2)(-q^\mu q^\nu+q^2\delta^{\mu\nu}),\\
j^\mu(x)&=&\sum_{i} Q_i\bar\psi_i(x)\gamma^\mu\psi_i(x).
\end{eqnarray}
$j^\mu(x)$ is the electromagnetic current, and $Q_i$ is the quark electric charge in units of the electron charge $e$ (the sum is over active flavors).
The form in the second equation is dictated by Lorentz and gauge symmetries. 

In the following it is convenient to use the time-momentum representation~\cite{Bernecker:2011gh} which results from interchanging the order of the Fourier transform and momentum integrals in Eqs.~(\ref{eq:ft}) and~(\ref{eq:amu}), respectively.
\begin{eqnarray}
\label{eq:hvp}
\Pi(q^2) -\Pi(0) &=& \sum_t \left(
\frac{\cos{qt} -1}{q^2} +\frac{1}{2} t^2
\right) C(t),\\
\label{eq:corr}
C(t) &=& \frac{1}{3}\sum_{\vec x,i}\langle j^i(\vec x, t)j^i(0)\rangle,\\
\label{eq:kernel}
 w(t) &=& 4 \alpha^2 \int_{0}^{\infty} {d \omega^2} f(\omega^2)\left[\frac{\cos{\omega t} -1}{\omega^2} +\frac{t^2}{2}\right],
 \end{eqnarray}
 where $C(t)$ is the Euclidean time correlation function, averaged over spatial directions, and Eq.~(\ref{eq:amu}) becomes
 \begin{eqnarray}
 \label{eq:t-m amu}
a_{\mu}^{\rm HVP}(T) &=& \sum_{{t=-T/2}}^{T/2}  w(t) C(t)=2\sum_{t=0}^{T/2}  w(t) C(t).
\end{eqnarray}
$T$ is the temporal size of the lattice, and $a_{\mu}^{\rm HVP}$ is obtained in the limit $T\to \infty$.  We have anticipated the use of the lattice with a discrete version of Eq.~(\ref{eq:t-m amu}). The weight $w(t)$ is sometimes modified by replacing the continuum Euclidean momentum-squared with its lattice version~\cite{Blum:2018mom}:
\begin{equation}
\label{eq:what} 
\hat{w}(t) = 4 \alpha^2 \int_{0}^{\infty} {d \omega^2} f(\omega^2)\left[\frac{\cos{\omega t} -1}{(2\sin{(\omega/2)})^2} +\frac{t^2}{2}\right].
\end{equation}
Note the double subtraction~\cite{Bernecker:2011gh,Lehner:2015bga,Aubin:2015rzx} in the cosine term in Eq.~(\ref{eq:hvp}):  $t^2/2$ cancels $\Pi(0)$ ``configuration-by-configuration" while the leading finite size correction is killed by the ``-1". The latter arises since $\Pi_{\mu\nu}(q^2)$ does not vanish as $q^2\to0$ when the time extent of the lattice is finite~\cite{Bernecker:2011gh}, but instead leads to a thermal electric susceptibility. In fact such terms are not constrained by the Ward--Takahashi Identity which in infinite volume leads to Eq.~(\ref{eq:wi}) and are allowed by the lattice symmetries~\cite{Bernecker:2011gh,Aubin:2015rzx}.
\subsection{Finite volume chiral perturbation theory}
\label{sec:fv chipt}
In this section, we consider the calculation of finite-volume effects in $a_\mu^{\rm HVP}$ to two loops, 
or next-to-next-to-leading order (NNLO) in chiral perturbation theory (ChPT), with the aim of
correcting our lattice result for $a_\mu^{\rm HVP}$ for finite-volume effects. 
With our pion masses near the physical value, it is safe
to assume that even at NNLO the most significant finite-volume correction will come from pion
loops, and we can thus restrict our calculation to isospin-symmetric two-flavor ChPT.

  There are two possible
strategies for doing this.   One is to first carry out a continuum extrapolation, and using results from continuum ChPT to correct for finite-volume effects.   The other is to correct the results at each lattice spacing, to obtain infinite-volume results at fixed lattice spacing.   As we are using
staggered fermions, the second strategy requires the use of staggered ChPT (SChPT)
\cite{Lee:1999zxa,Aubin:2003mg}.  If all our ensembles
were at the same pion mass and volume, the two methods should yield equivalent results.   However, both the pion masses and volumes of the three ensembles are slightly different ({\it cf.} Table \ref{tab:ensembles}).   In this
case, applying the finite-volume correction at a fixed lattice spacing has the advantage that this automatically corrects for the slightly different volumes.\footnote{But not the slightly different pion masses \cite{GMS}.}   While a full two-loop SChPT calculation is outside the scope of this paper, it is easy to
change the NLO continuum ChPT result into a SChPT result; one only has to carry out a weighted average over the different taste pion masses for a given ensemble \cite{Aubin:2015rzx}.   In practice, what we will
do is to first correct the finite-volume lattice results for $a_\mu^{\rm HVP}$ using NLO SChPT, then
extrapolate to the continuum limit, after which we apply the remaining NNLO continuum ChPT correction.
Because of the slight mistunings of the pion masses and volumes, there will be a systematic error
associated with this last step, but this systematic error will be much smaller than it would be
if we were to extrapolate to the continuum first, and then apply NLO plus NNLO continuum ChPT to
correct for finite-volume effects.

While the vacuum polarization in finite volume to two loops has been calculated before in momentum space \cite{BR},\footnote{For recent work on finite-volume effects of order $\mbox{exp}[-m_\pi L]$ not using ChPT, see Ref.~\cite{Hansen:2019rbh}.}
we will directly {carry out the ChPT calculation of} $C(t)$, defined in {Eq.~(\ref{eq:corr})}, in the time-momentum representation, for
$t>0$, in a spatial volume of linear size $L$, with periodic boundary conditions.\footnote{We take
the time extent to be infinite.}   This makes the calculation somewhat simpler, because we do not have to consider
diagrams that lead to contributions proportional to $\delta(t)$ (which, in momentum space,
correspond to contact terms).   Our result will depend on only two low-energy constants,
$F$, the pion decay constant in the chiral limit, and $\ell_6$, which is an order-$p^4$
low-energy constant appearing in the EM current at this order.\footnote{We use the notation and
conventions of Ref.~\cite{GLAP} for low-energy constants.}

Of course, the ChPT expression for $C(t)$ is only reliable for large $t$, while $C(t)$ for all $t>0$
is needed in the sum~(8).\footnote{$C(0)$ is not needed as the weight $w(t)\propto t^4$ for 
small $t$.}   However, as already observed in Ref.~\cite{Aubin:2015rzx}, finite-volume effects are a long-distance
effect, and one thus expects the finite-volume correction to this correlation function to be 
reliably estimated for all $t>0$, so that we can, in fact, estimate the
finite-volume effect in $a_\mu^{\rm HVP}$ using ChPT.   An advantage is that this avoids using
models to go beyond NLO ChPT (which is the same as scalar QED), as was proposed in 
Ref.~\cite{mainz13}.   As we will see, the ChPT result for the difference 
\begin{equation}
\label{deltaamu}
\Delta a_\mu^{\rm HVP} = \lim_{L\to\infty}a_\mu^{\rm HVP}(L)- a_\mu^{\rm HVP}(L)
\end{equation}
is indeed well defined.\footnote{ In general we define $\Delta f(L)= \lim_{L\to\infty}f(L)-f(L)$} in what follows.

The pion contribution to the EM current, to the order we need, is given by\footnote{There are contributions from other order-$p^4$ low-energy constants, but they do not appear in the result for $C(t)$ after mass renormalization.}
\begin{eqnarray}
\label{EMcurrent}
j_\mu(x)&=&i\left(\pi^-\partial_\mu\pi^+-\pi^+\partial_\mu\pi^-\right)\left(1-\frac{1}{3F^2}
\left((\pi^0)^2+2\pi^+\pi^-\right)\right)\\
&&-\frac{2i\ell_6}{F^2}\,\partial_\nu\left(\partial_\mu\pi^+\partial_\nu\pi^--
\partial_\nu\pi^+\partial_\mu\pi^-\right)\ .\nonumber
\end{eqnarray}
   Working in Euclidean space, a relatively
straightforward calculation in the time-momentum representation yields the result for $C(t)$ to NNLO 
{in the continuum limit} as
\begin{eqnarray}
\label{CtNNLOdimreg}
C(t)&=&\frac{10}{9}\frac{1}{3}\Biggl(\frac{1}{L^d}\sum_{{\vec p}}\frac{{\vec p}^2}{E_p^2}\,e^{-2E_p t}\Biggl[1-\frac{2}{F^2}D(m_\pi^2)-\frac{8({\vec p}^2+m_\pi^2)}{F^2}\,\ell_6\Biggr]\\
&&\phantom{\frac{10}{9}\frac{1}{3}\,e^2\Biggl(}+\frac{1}{2dF^2}\frac{1}{L^{2d}}\sum_{{\vec p},{\vec k}}
\frac{{\vec p}^2{\vec k}^2}{E_p^2E^2_k}\frac{E_k e^{-2E_p t}-E_p e^{-2E_k t}}{{\vec k}^2-{\vec p}^2}
\Biggr)\ ,
\nonumber
\end{eqnarray}
in which
\begin{eqnarray}
\label{energy}
E_p&=&\sqrt{m_\pi^2+{\vec p}^2}\ ,\\
D(m_\pi^2)&=&\frac{1}{L^d}\sum_{\vec k}\frac{1}{2E_k}\ ,
\end{eqnarray}
and the sums over $\vec p$ and $\vec k$ are over the momenta $2\pi{\vec n}/L$, $n_i$ integer, 
in a box with periodic boundary conditions.   In Eq.~(\ref{CtNNLOdimreg}) we gave the result in
$d=3+\epsilon$ spatial dimensions in order to regulate the UV divergence present in $d=3$.
After defining a renormalized $\ell_6^r$ by
\begin{equation}
\label{l6ren}
\ell_6=\ell_6^r(\mu)-\frac{1}{3}\,\frac{1}{16\pi^2}\left(\frac{1}{\epsilon}-\log\mu-\frac{1}{2}(\log{(4\pi)}-\gamma+1)\right)\ ,
\end{equation}
the limit $d\to 3$ can be taken, yielding a finite result for $C(t)$.   The factor $10/9$ is needed
to isolate the light-quark connected part \cite{Aubin:2015rzx,mainz13,DellaMorte:2010aq}.   The pion mass $m_\pi$
appearing in Eq.~(\ref{CtNNLOdimreg}) is the renormalized (physical) pion mass.   This
renormalization absorbs the low-energy constants $\ell_{3,4}$ which appear in the explicit
calculation.   We note that the terms in the double sum on the second line of Eq.~(\ref{CtNNLOdimreg})
with ${\vec p}^2={\vec k}^2$ lead to a term proportional to $t\,e^{-2E_p t}$, leading to the
expected energy shift for two pions in an $I=1$, $\ell=1$ state in a finite volume \cite{ML}.

In order to extract the finite-volume corrections, we use the Poisson resummation formula
\begin{equation}
\label{Presum}
\sum_{\vec n}\delta^{(d)}\left({\vec p}-\frac{2\pi{\vec n}}{L}\right)
=\sum_{\vec n}\frac{L^d}{(2\pi)^d}\,\delta^{(d)}\left(\frac{L{\vec p}}{2\pi}-{\vec n}\right)
=\frac{L^d}{(2\pi)^d}\,\sum_{\vec n}e^{i{\vec n}\cdot{\vec p}L}\ .
\end{equation}
Let us work out the extraction of $\Delta a_\mu^{\rm HVP}$ to NLO in ChPT, relegating the treatment
of the 
NNLO contribution to the appendix.   The NLO part of $C(t)$ is obtained by dropping all terms
of order $1/F^2$ in Eq.~(\ref{CtNNLOdimreg}).  Employing Eq.~(\ref{Presum}), the NLO part
$C^{\rm NLO}(t)$ can be written as
\begin{equation}
\label{CtNLOresum}
C^{\rm NLO}(t)=-\frac{10}{9}\frac{1}{6\pi^2}\sum_{n^2=0}^\infty Z_{00}(0,n^2)\frac{1}{nL}
\int_0^\infty dp\,\frac{p^3}{E_p^2}\,e^{-2E_p t}\,\sin{(npL)}\ ,
\end{equation}
where $n^2$ is summed over all non-negative integers and \cite{ML}
\begin{equation}
\label{Zeta00}
Z_{00}(0,{\vec n}^2)=-\sum_{{\vec m},{\vec m}^2={\vec n}^2}\,1\ .
\end{equation}
The $n^2=0$ term, with $\sin{(npL)}/(nL)\to p$ and $Z_{00}(0,0)=-1$, yields $C^{\rm NLO}(t)$ in the infinite-volume limit.
Inserting Eq.~(\ref{CtNLOresum}) into Eq.~(\ref{eq:t-m amu}) with $T=\infty$ and replacing the sum over $t$ by an integral,
we find that
\begin{equation}
\label{amuNLO}
\Delta a_\mu^{\rm HVP,\,NLO}=\frac{10}{9}\frac{\alpha^2}{6\pi^2} \sum_{n^2=1}^\infty
 \frac{Z_{00}(0,n^2)}{nL}\int_0^\infty dp\,\frac{p^3}{E_p^2}\,\sin{(npL)}F(p^2)\ ,
 \end{equation}
with
\begin{eqnarray}
\label{Fdef}
F(p^2)&=&\int_0^\infty dq^2 f(q^2)\frac{q^2}{E_p^3(4E_p^2+q^2)}\\
&=&-\frac{8 E_p^2 -m_\mu^2}{2 E_p^3
   m_\mu^2 } +\frac{8(2
   E_p^2-m_\mu^2)}{ E_p
   m_\mu^4 }   \log \left(\frac{2
   E_p}{m_\mu}\right)\nonumber\\
&&+\frac{\left(8 E_p^4-8 E_p^2
   m_\mu^2+ m_\mu^4\right)}{ E_p^2
   m_\mu^4 \sqrt{ E_p^2-
   m_\mu^2}}  
    \log
   \left(\frac{-2 E_p \sqrt{
   E_p^2- m_\mu^2}+2
   E_p^2- m_\mu^2}{
   m_\mu^2}\right)\ .\nonumber
\end{eqnarray}
Using the parameter values 
of Table I, we then obtain
\begin{equation}
\label{DamuNLOnum}
\Delta a_\mu^{\rm HVP,\,NLO}=\left\{\begin{array}{c} 20.59\times 10^{-10}\ ,\quad L/a=96\\
21.60\times 10^{-10}\ ,\quad L/a=64\\
18.08\times 10^{-10}\ ,\quad L/a=48
\end{array}\right.\ .
\end{equation}
Adding the NNLO contributions computed in the appendix and given in Eq.~(\ref{DamuNNLOnumtotal}), we find for the total finite-volume
correction
\begin{equation}
\label{Damunum}\Delta a_\mu^{\rm HVP}=\left\{\begin{array}{c} (29.7\pm 4.0)\times 10^{-10}\ ,\quad L/a=96\\
(30.6\pm 3.8)\times 10^{-10}\ ,\quad L/a=64\\
(25.5\pm 3.0)\times 10^{-10}\ ,\quad L/a=48
\end{array}\right.\ .
\end{equation}
The errors are estimated as follows.   The NNLO contribution is of order $0.4$--$0.45$ times
the NLO contribution.   We then assume that the next order in ChPT, which we did not 
compute, is again of order $0.4$--$0.45$ times the NNLO contribution, and we use this 
estimate as our error.

The fact that the three values in Eq.~(\ref{Damunum}) are different is due to the mistuning of the
pion masses and volumes of the three ensembles.  If we were to apply the correction to the
continuum extrapolated value of $a_\mu^{\rm HVP}$, we would thus have to use some average,
and the spread of $5.1\times 10^{-10}$ between the three values would represent a systematic
error associated with the mistuning.  If we were to apply only the NNLO correction in the 
continuum limit, that spread would be reduced to $1.7\times 10^{-10}$ ({\it cf.} Eq.~(\ref{DamuNNLOnumtotal})).   Hence, as explained
above, what we will do is to first use NLO SChPT to correct the value of $a_\mu^{\rm HVP}$
at each lattice spacing, then extrapolate, and finally apply the NNLO correction computed
in Eq.~(\ref{DamuNNLOnumtotal}) in the Appendix.   

{In order to adapt the NLO result~(\ref{amuNLO}) to the staggered case, all that needs to be done
is to average $C^{\rm NLO}(t)$ of Eq.~(\ref{CtNLOresum}) over the taste-split pion spectrum
$m_\pi=m_P$, $m_A$, $m_T$, $m_V$ and $m_I$, with weights $1/16$, $1/4$, $3/8$, $1/4$ and $1/16$,
respectively.}
Using the taste-split pion spectrum for each ensemble,\footnote{We thank Doug Toussaint for
providing the pion spectra, and for discussions of the taste splittings.} we find for the staggered
NLO {finite-volume} corrections for each ensemble the values 
\begin{equation}
\label{stagDamunum}
\Delta a_\mu^{\rm HVP}=\left\{\begin{array}{c} 15.6\times 10^{-10}\ ,\quad L/a=96\\
~6.9\times 10^{-10}\ ,\quad L/a=64\\
~2.1\times 10^{-10}\ ,\quad L/a=48
\end{array}\right.\ .
\end{equation}

Finally, the $n^2=0$ term in Eq.~(\ref{CtNLOresum}) gives us access to the effect of taste breaking in the pion masses in infinite volume, to NLO in ChPT.  We use this to compute the corresponding corrections for each of
our ensembles, finding these to be equal to 
\begin{equation}
\label{tastecorrection}
\Delta_{\rm taste} a_\mu^{\rm HVP}=\left\{\begin{array}{c} ~9.5\times 10^{-10}\ ,\quad L/a=96\\
34.2\times 10^{-10}\ ,\quad L/a=64\\
51.6\times 10^{-10}\ ,\quad L/a=48
\end{array}\right.\ .
\end{equation}
These corrections are to be added to the lattice result to correct for taste breaking in the pion spectrum in infinite volume, to NLO in ChPT.
Of course, since taste breaking is a lattice-spacing effect, whether one adds these corrections or not should not matter in the continuum limit.   The 
difference one finds between values extrapolated to the continuum limit with or without this correction thus
gives an estimate of the systematic error associated with taking the continuum limit.
{Adding both Eq.~(\ref{stagDamunum}) and Eq.~(\ref{tastecorrection}) to the numerical lattice results will correct, at NLO, for finite-volume effects
(Eq.~(\ref{stagDamunum})) and taste-breaking effects (Eq.~(\ref{tastecorrection})).  Lattice results corrected only with Eq.~(\ref{stagDamunum}) will be shown in third column of Table~III below, while those corrected with both Eqs.~(\ref{stagDamunum}) and (\ref{tastecorrection})
will be shown in the 4th column.   As already stated above, NNLO finite-volume corrections will only be applied after the continuum
limit has been taken.}

\subsection{Lattice details}

The computation rests heavily on the use of noise reduction techniques developed by the RBC and UKQCD collaborations, including all-mode (AM) and full volume low-mode (LM) averaging (see Refs.~\cite{Blum:2012uh,Shintani:2014vja,Giusti:2004yp,DeGrand:2004qw,Neff:2001zr,Giusti:2005sx,Blum:2018mom}). 

We take a moment to describe the low-mode structure of the staggered fermion Dirac operator which plays a central role. For valence quarks we use the highly improved staggered quarks (HISQ)~\cite{Follana:2006rc} fermion Dirac operator minus the Naik term, so the following, which is true in general for naive staggered fermions, applies here. The staggered Dirac operator is the sum of a hermitian mass term which commutes with an anti-hermitian hopping term, so it satisfies (using even-odd ordering of sites)
\begin{eqnarray}
M\left(
\begin{array}{c}
n_o \\ n_e
\end{array}
\right) 
=\left(
\begin{array}{cc}
 m  & M_{oe}\\
 M_{eo} &  m
\end{array}
\right)
\left(
\begin{array}{c}
n_o \\ n_e
\end{array}
\right) &=&
(m + i\lambda_n) \left(
\begin{array}{c}
n_o \\ n_e
\end{array}
\right), 
\end{eqnarray}
where $m$ is the quark mass and $M_{oe}$ hops quarks from even to odd sites. Similarly, the preconditioned operator $M^\dagger M$ which is used in practice satisfies
\begin{eqnarray}
\left(
\begin{array}{cc}
 m  & -M_{oe}\\
 -M_{eo} &  m
\end{array}
\right)
\left(
\begin{array}{cc}
 m  & M_{oe}\\
 M_{eo} &  m
\end{array}
\right)
\left(
\begin{array}{c}
n_o \\ n_e
\end{array}
\right)
&=& \nonumber\\
\left(
\begin{array}{cc}
m^2  - M_{oe} M_{eo} & 0\\
0 & m^2  - M_{eo} M_{oe}
\end{array}
\right)
\left(
\begin{array}{c}
n_o \\ n_e
\end{array}
\right) &=&
(m^2+\lambda_n^2)
\left(
\begin{array}{c}
n_o \\  n_e
\end{array}
\right).
\end{eqnarray}
Eigenvectors of the preconditioned operator are eigenvectors of $M$ with squared magnitude eigenvalue, and the even part can be obtained from the odd part, 
\begin{equation}
n_e =\frac{ -i}{\lambda_n} M_{eo} n_o.    
\end{equation}
The eigenvalues come in $\pm $ pairs: If $n_+=(n_o, n_e)$ is an eigenvector with eigenvalue $\lambda_n$, then $n_- = (-n_o,n_e)$ is also an eigenvector with eigenvalue $-\lambda_n$:
\begin{eqnarray}
\left(
\begin{array}{cc}
m  & M_{oe}\\
 M_{eo} &  m
\end{array}
\right)
\left(
\begin{array}{c}
-n_o \\ n_e
\end{array}
\right) &=&
(m - i\lambda_n) \left(
\begin{array}{c}
-n_o \\ n_e
\end{array}
\right).
\label{eq:minus lambda evec}
\end{eqnarray}
Thus we can construct pairs of eigenvectors, $n_+$, $n_-$, corresponding to $\pm i \lambda$ for each eigen-pair ($\lambda^2$, $n_o$) computed with the Lanczos algorithm.

The full-volume LMA takes advantage of the spectral decomposition of the quark propagator that requires only two independent volume sums instead of a volume-squared sum in the correlation function. We employ a conserved current (again, minus the three-hop Naik term) which makes the ``meson fields" a bit more complicated, 
\begin{eqnarray}
J^\mu(x) &=&  -\frac{1}{2} \eta_\mu(x) \left(
\bar{\chi}(x+\hat{\mu}) U^\dagger_\mu(x)\chi(x)
~+~\bar{\chi}(x)U_\mu(x)\chi(x+\hat{\mu})\right).
\end{eqnarray}
$\chi(x)$ are single component staggered fermion fields whose spinor nature is encoded in the staggered phases, $\eta(x)$, arising from the spin diagonalization of the fermion action. The gauge links $U_\mu(x)$ ensure the point-split current is gauge invariant.  A spectral decomposition of the low-mode part of the quark propagator is used in the AMA and LMA procedures,
\begin{eqnarray}
M^{-1}_{x,y} &=& \sum_n^{N_{\rm low}}
\left(\frac{\langle x |n_+\rangle\langle n_+|y\rangle}{m+i\lambda_n}+
\frac{\langle x |n_-\rangle\langle n_-|y\rangle}{m-i\lambda_n}\right)\ ,\nonumber
\end{eqnarray}
where $N_{\rm low}$ is the number of low modes. The two point, current-current correlation function then becomes
{\begin{eqnarray}
4\sum_{{\vec x},\vec{y}} \langle J_\mu(t_x,\vec{x})J_\nu(t_y,\vec{y})\rangle&=&-\sum_{m,n}\sum_{{\vec x},\vec{y}}
\frac{1}{\lambda_m\lambda_n}\Biggl(\Lambda^\dagger_\mu(x)_{mn}\Lambda^\dagger_\nu(y)_{nm}+\Lambda^\dagger_\mu(x)_{mn}\Lambda_\nu(y)_{nm}\nonumber\\
&&\phantom{-\sum_{m,n}\sum_{{\vec x},\vec{y}}
\frac{1}{\lambda_m\lambda_n}}+\Lambda_\mu(x)_{mn}\Lambda^\dagger_\nu(y)_{nm}+\Lambda_\mu(x)_{mn}\Lambda_\nu(y)_{nm}\Biggr)\ ,\nonumber
\end{eqnarray}
}
where $\lambda_n$ is shorthand for either $m\pm i\lambda_{n}$, and the sums over eigenvectors run up to $2 N_{\rm low}$.  To compute the above we construct meson fields,
\begin{eqnarray}
(\Lambda_{\mu}(t))_{n,m}
&=& \sum_{\vec x} {\langle n|x\rangle \eta_\mu(x) U_\mu(x)\langle x+\mu|m\rangle} (-1)^{(m+n)x+m},
\end{eqnarray}
with eigenvector ordering $\lambda_0,-\lambda_0, \lambda_1, -\lambda_1, \dots, \lambda_{N_{\rm low}}, -\lambda_{N_{\rm low}}$.
The factor $(-1)^{(m+n)x+m}$ arises from the construction of $n_-$ from $n_+$ since even $m$ or $n$ always corresponds to $n_+$ while odd corresponds to $n_-$. 

The AMA and LMA procedures are used to produce an improved estimator for the expectation value of any observable $O$ by adding and subtracting terms that are exactly equal in the infinite statistics limit. Outside this limit the unimproved and improved estimates are statistically equivalent, with the latter having smaller errors { (assuming the same computational expense)}. 
The combined AMA and full-volume LMA improved estimator is given by
\begin{eqnarray}
\label{eq:ama+lma}
\langle O\rangle &=& \langle O\rangle_{\rm exact} - \langle O\rangle_{\rm approx} + \frac{1}{N} \sum_{i} \langle O_i\rangle_{\rm approx}- \frac{1}{N} \sum_{i} \langle O_i\rangle_{\rm LM} 
+ \frac{1}{V} \sum_{i} \langle O_i\rangle_{\rm LM}.
\end{eqnarray}
The first three terms on the right hand side of Eq.~(\ref{eq:ama+lma}) correspond to AMA~\cite{Blum:2012uh,Shintani:2014vja} while the last two supplement this with the full-volume LMA~\cite{Giusti:2004yp,DeGrand:2004qw,Neff:2001zr,Giusti:2005sx,Blum:2018mom}. The expensive ``exact" (to numerical precision) calculation is done relatively seldom while the inexpensive ``approx" calculation is done often to reduce the statistical error. The difference of the first term with the 2nd and 4th terms corrects the bias induced by the 3rd and 5th approximate terms. Note that in this work the first two sums in Eq.~(\ref{eq:ama+lma}) are taken over a uniform grid of point-source propagators on a time slice (see Tab.~\ref{tab:ensembles}) which is much smaller in number than the total number of lattice sites summed over for the final sum in Eq.~(\ref{eq:ama+lma}). The approximate propagators are computed with a relaxed conjugate gradient stopping residual, $10^{-5}$, while the exact is set to $10^{-8}$. Both are deflated, that is a number of exact low-modes of the Dirac operator are used to compute each (see Tab.~\ref{tab:ensembles}).
\section{results}
\label{sec:results}

We use the 2+1+1 flavor, physical mass ensembles generated by the MILC collaboration at three lattice spacings shown in Tab.~\ref{tab:ensembles}. They have roughly the same physical extent, $L \sim 5.5-5.8$ fm.
\begin{table}[htp]
\begin{center}
\begin{tabular}{|c|c|c|c|c|c|c|c|}
\hline
 &  &  &  &  &  & AMA & measurements \\
$m_\pi$ (MeV) & $a$ (fm) & size & $L$ (fm) & $m_\pi L$ & LM & srcs &(approx-exact-LMA) \\
\hline
133 & 0.12121(64) & $48^3\times 64$ & 5.82 & 3.91 & 3000 & $4^3\times4$ &26-26-26 \\
130 & 0.08787(46) &  $64^3\times 96 $ & 5.62 & 3.66 & 3000 & $4^3\times4$ &36-36-40 \\
134 & 0.05684(30) &  $96^3\times 192$& 5.46 & 3.73 & 2000 &$3^3\times8$&22-22-23\\\hline
\end{tabular}
\end{center}
\caption{Gauge field ensemble parameters  \cite{Bazavov:2014wgs}. ``LM" is the number of low-modes of the preconditioned Dirac operator. ``AMA srcs" is the number of approximate point source propagators on each configuration which are spread uniformly over several time slices. The number of exact point source propagators per configuration is eight for each ensemble. The number of configurations used for approximate, exact, and LMA measurements in this study are given in the last column.}
\label{tab:ensembles}
\end{table}%

\begin{figure}[htbp]
\begin{center}
\includegraphics[width=0.55\textwidth]{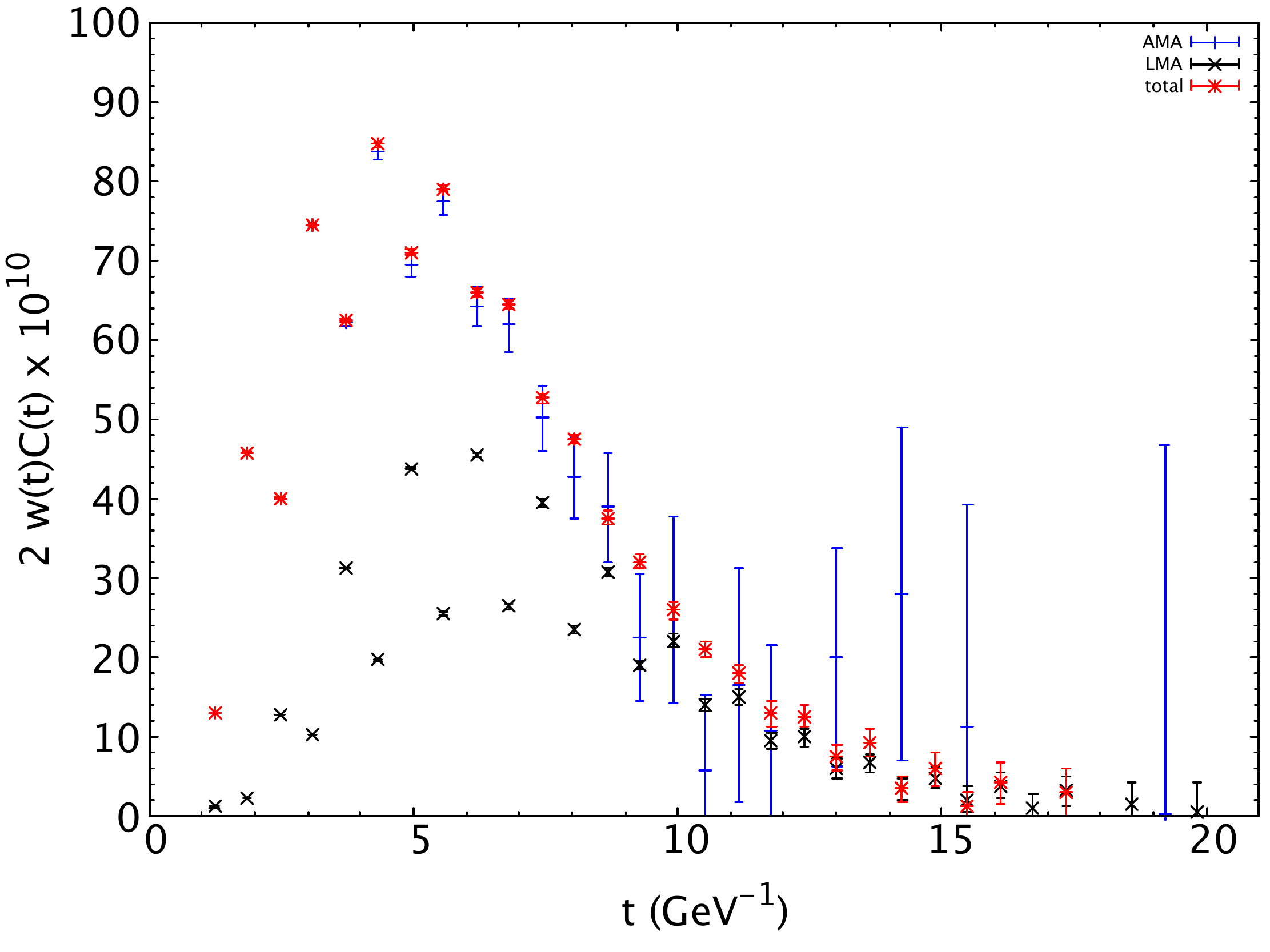}
\includegraphics[width=0.55\textwidth]{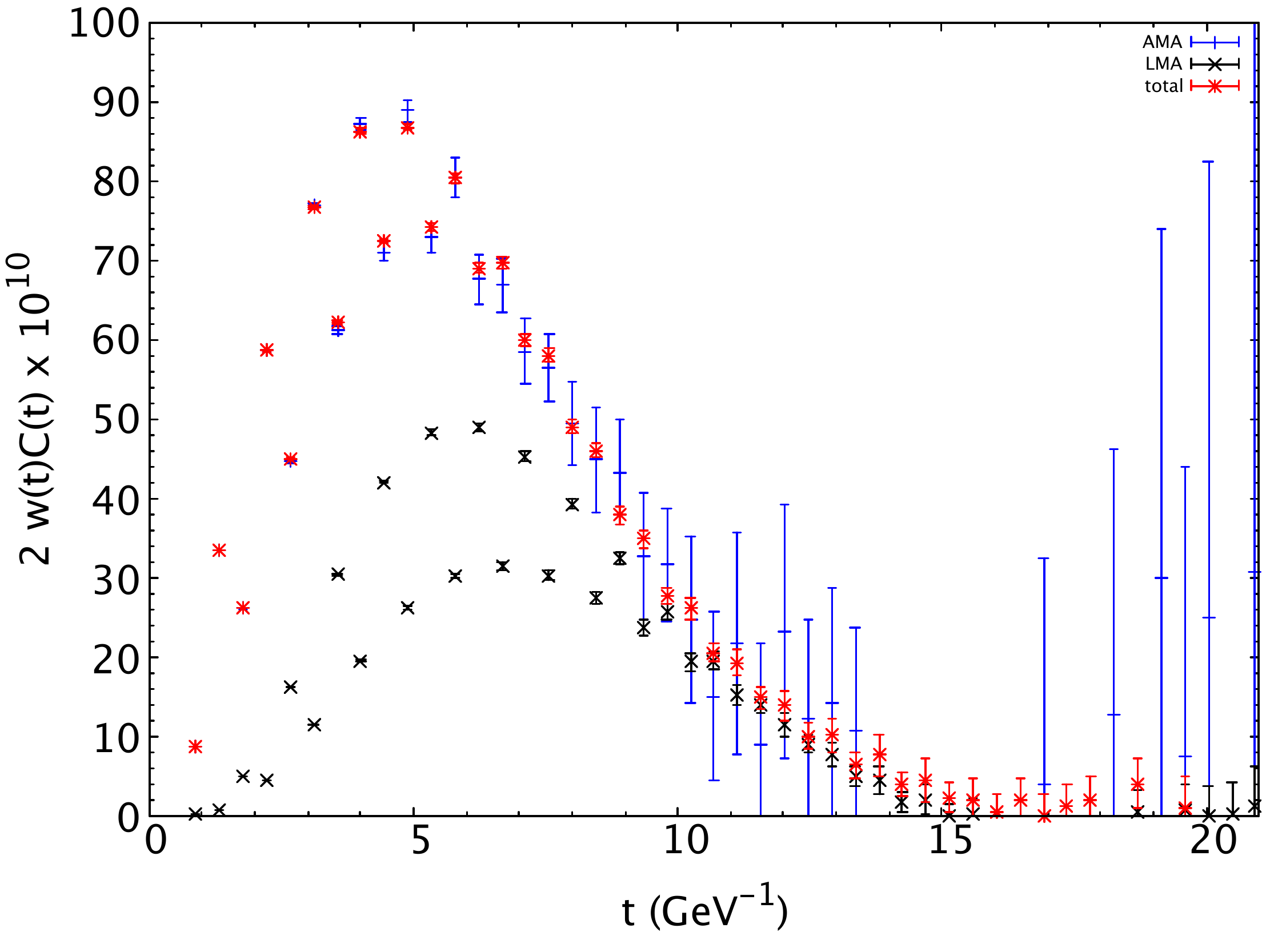}
\includegraphics[width=0.55\textwidth]{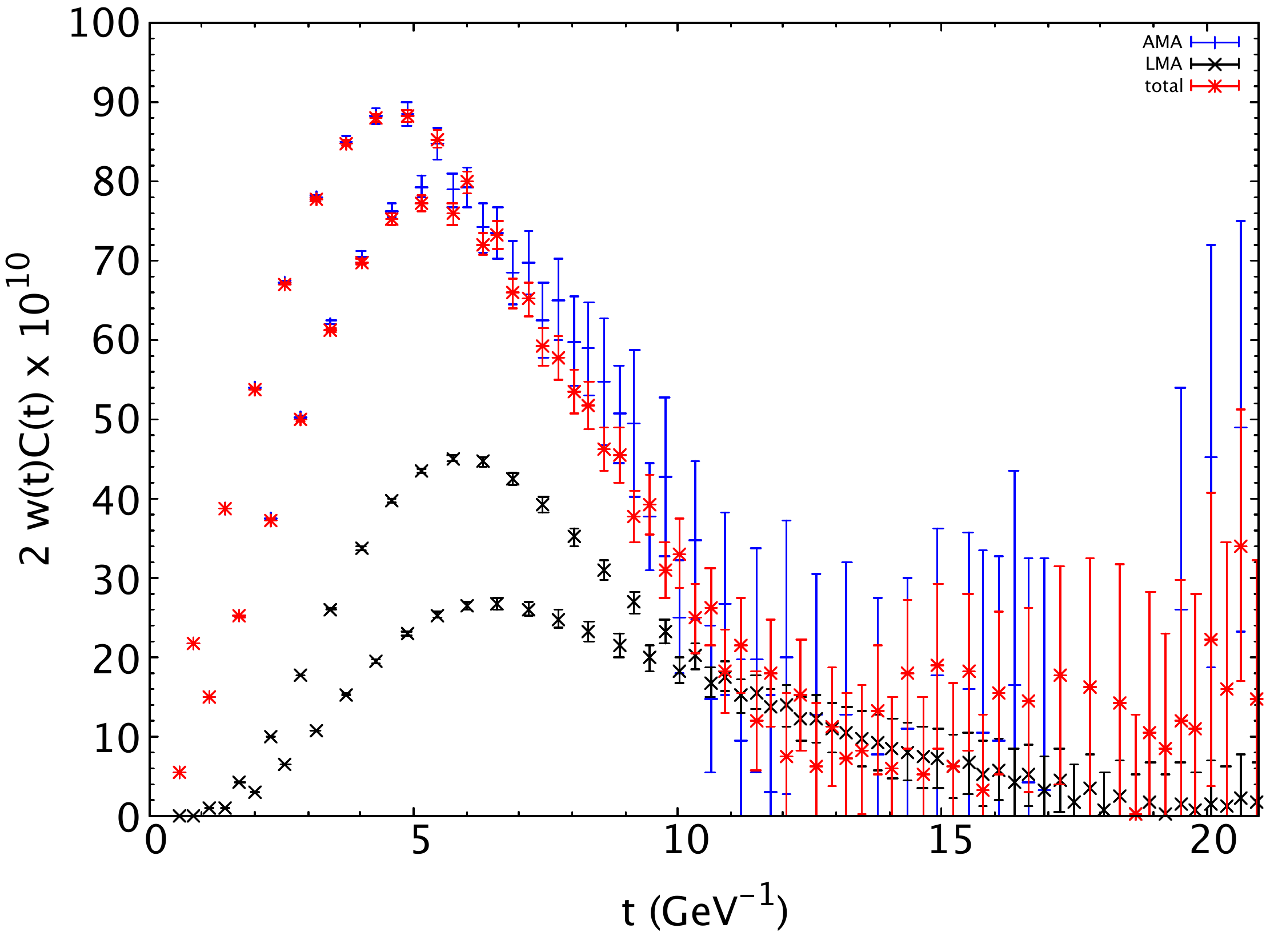}
\caption{The summand in Eq.~(\ref{eq:t-m amu}) for each ensemble in Tab.~\ref{tab:ensembles} (from top, coarsest to finest). Total ({red} stars) refers to the sum in Eq.~(\ref{eq:ama+lma}). Also shown are the low-mode ({black} crosses) and AMA ({blue} plusses) contributions. Odd-parity, excited state oscillations intrinsic to staggered fermions are readily apparent.}
\label{fig:amu integrand}
\end{center}
\end{figure}

In Fig.~\ref{fig:amu integrand} the summand in Eq.~(\ref{eq:t-m amu}) for each ensemble is shown along with the full volume LMA and AMA contributions. In the figure ``total" refers to the sum of five terms in Eq.~(\ref{eq:ama+lma}).
As observed in Ref.~\cite{Blum:2018mom} there is a huge reduction in statistical error from the low-mode average, the last term in Eq.~(\ref{eq:ama+lma}) (compare the total with full volume LMA and without, which is just AMA). The error reduction is especially large for large distance, as expected since the low-modes dominate this region. For the $96^3$ ensemble the number of low modes used was 2000 ($\times 2$) compared to the other two ensembles ($2\times 3000$), due to computer and memory resource limitations. This is unfortunate as one can see from Fig.~\ref{fig:amu integrand} that the full volume LMA is not as effective. Even though it appears that the low-mode contribution is mostly saturated (since it is comparable for all three ensembles), apparently the extra low-modes for the two coarser ensembles are very effective at reducing statistical noise.

\begin{figure}[htbp]
\begin{center}
\includegraphics[width=0.55\textwidth]{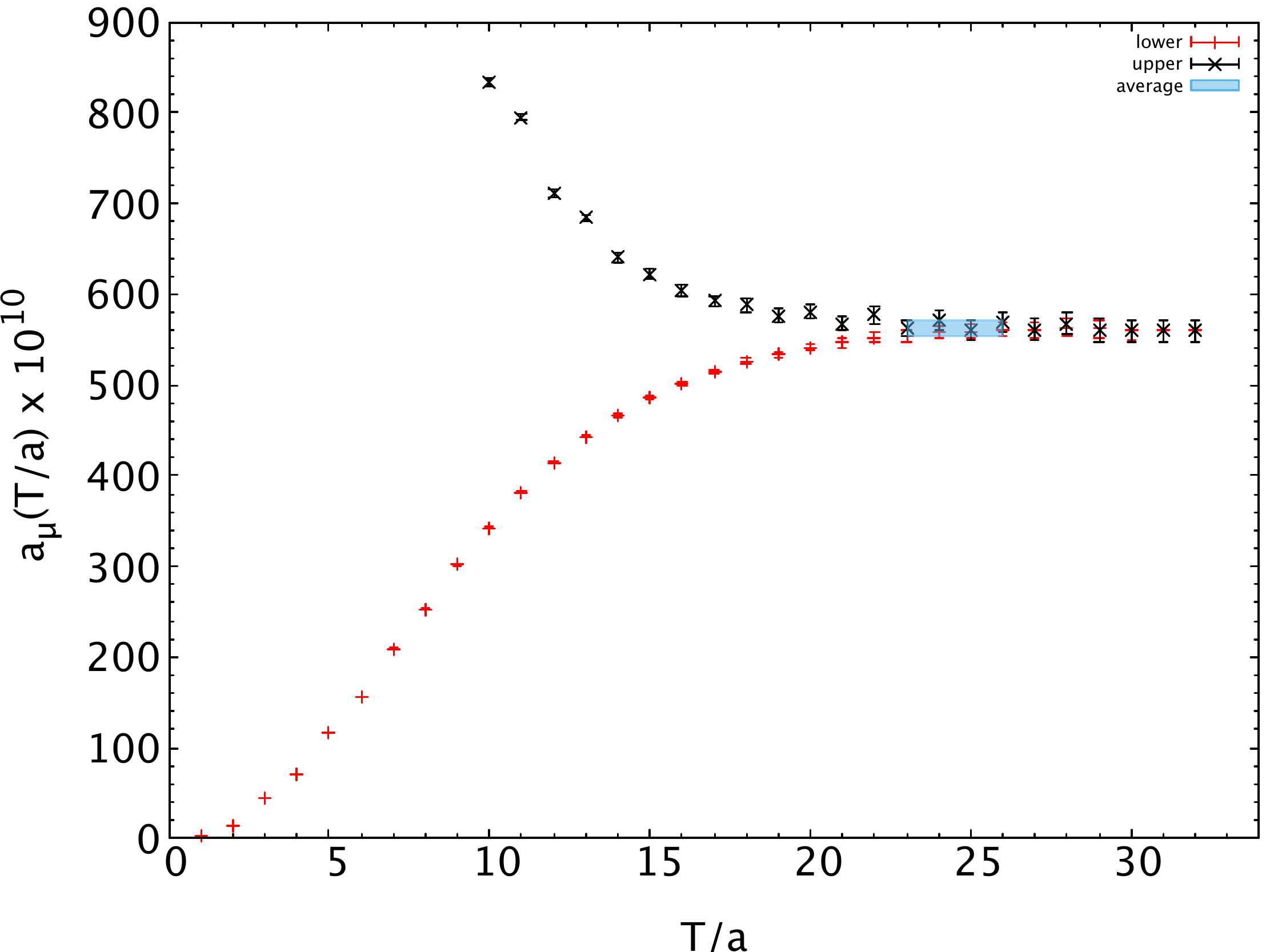}
\includegraphics[width=0.55\textwidth]{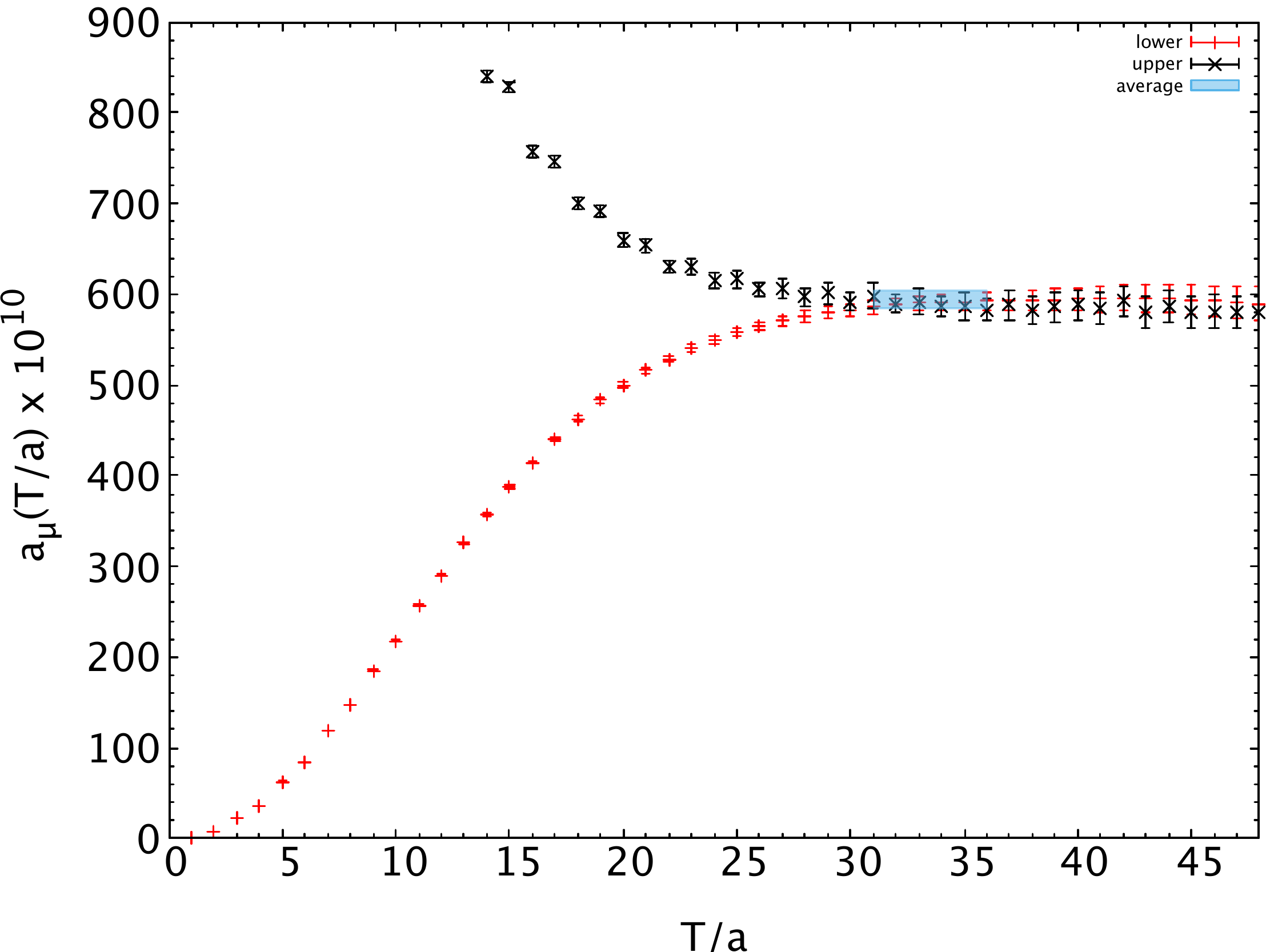}
\includegraphics[width=0.55\textwidth]{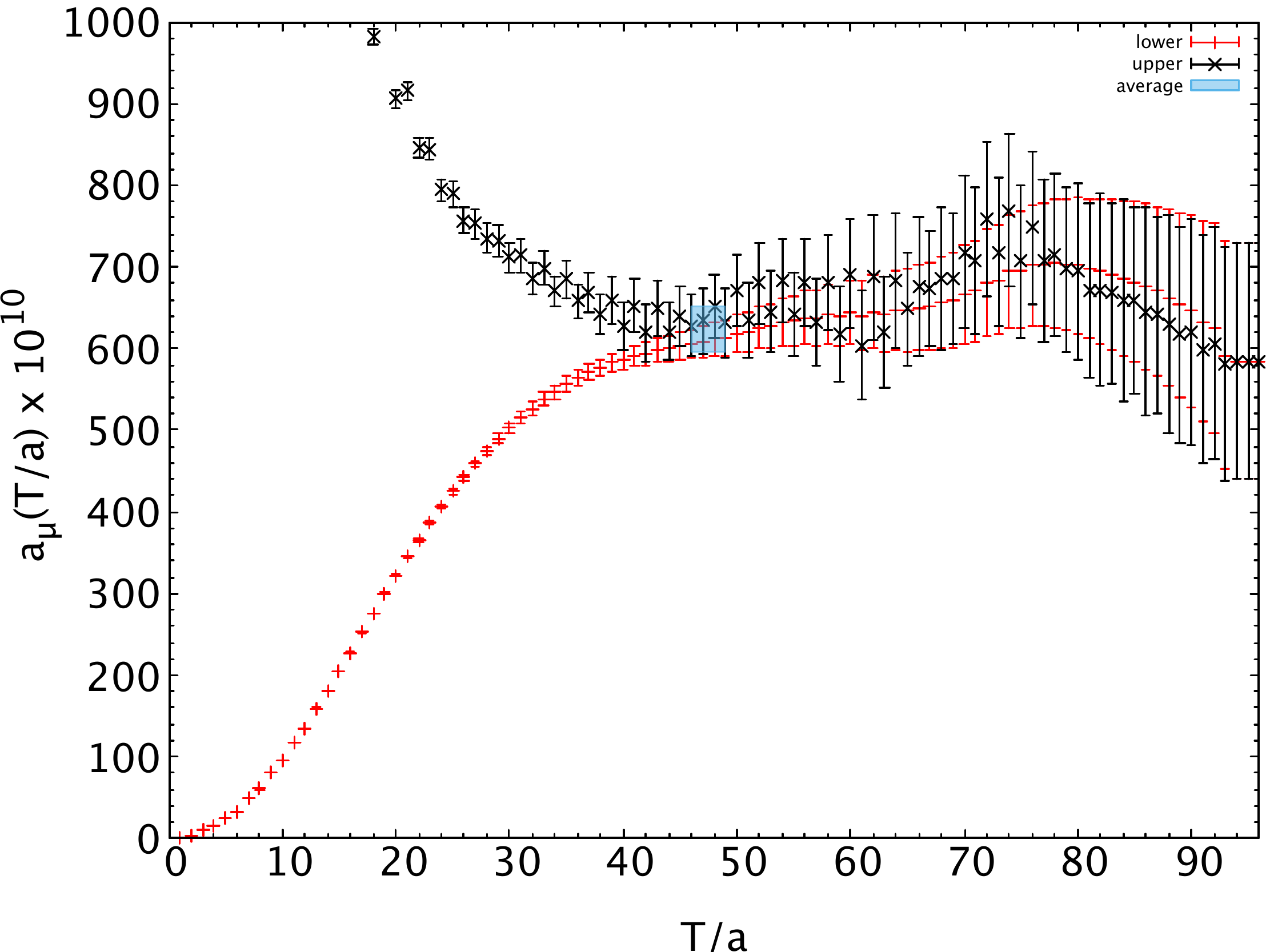}
\caption{Bounding method for total contribution to the muon anomaly, using the weighting function $w$. $48^3$ (top), $64^3$ (middle), and $96^3$ (bottom) ensembles. $T/a$ is the time slice where $C(t)$ switches over from the calculated value to the analytic value giving the upper {(black crosses)} or lower {(red plusses)} bound.{ The
blue shaded area indicates our averages.}}
\label{fig:bound}
\end{center}
\end{figure}

In order to reduce further the statistical errors on the integrated result, we employ the bounding method~\cite{Blum:2018mom,Borsanyi:2017zdw} wherein $C(t)$, for $t>T$, is given by $C(t)=0$ (lower bound), and $C(t)=C(T) e^{-E_0(t-T)}$ (upper bound), where $E_0=2\sqrt{m_\pi^2+(2\pi/L)^2}$, $i.e$, the lowest (two pion) energy state in the vector channel. At sufficiently large $T$ the bounds overlap, and an estimate for $a_\mu$ can be made which may be more precise than simply summing over the noisy long-distance tail. In Fig.~\ref{fig:bound} results are shown for each ensemble.  Central values for $a_\mu$ are averages over a suitable range where $T$ is large enough for the bounds to overlap but not so large that statistical errors blow up. We average the upper and lower bounds together over the ranges 2.7-3.2 fm for the $48^3$ and $64^3$ ensembles, and 2.6-2.8 fm for $96^3$. The statistical errors on the averages are computed using the jackknife method.

\begin{figure}[htbp]
\begin{center}
\includegraphics[width=0.7\textwidth]{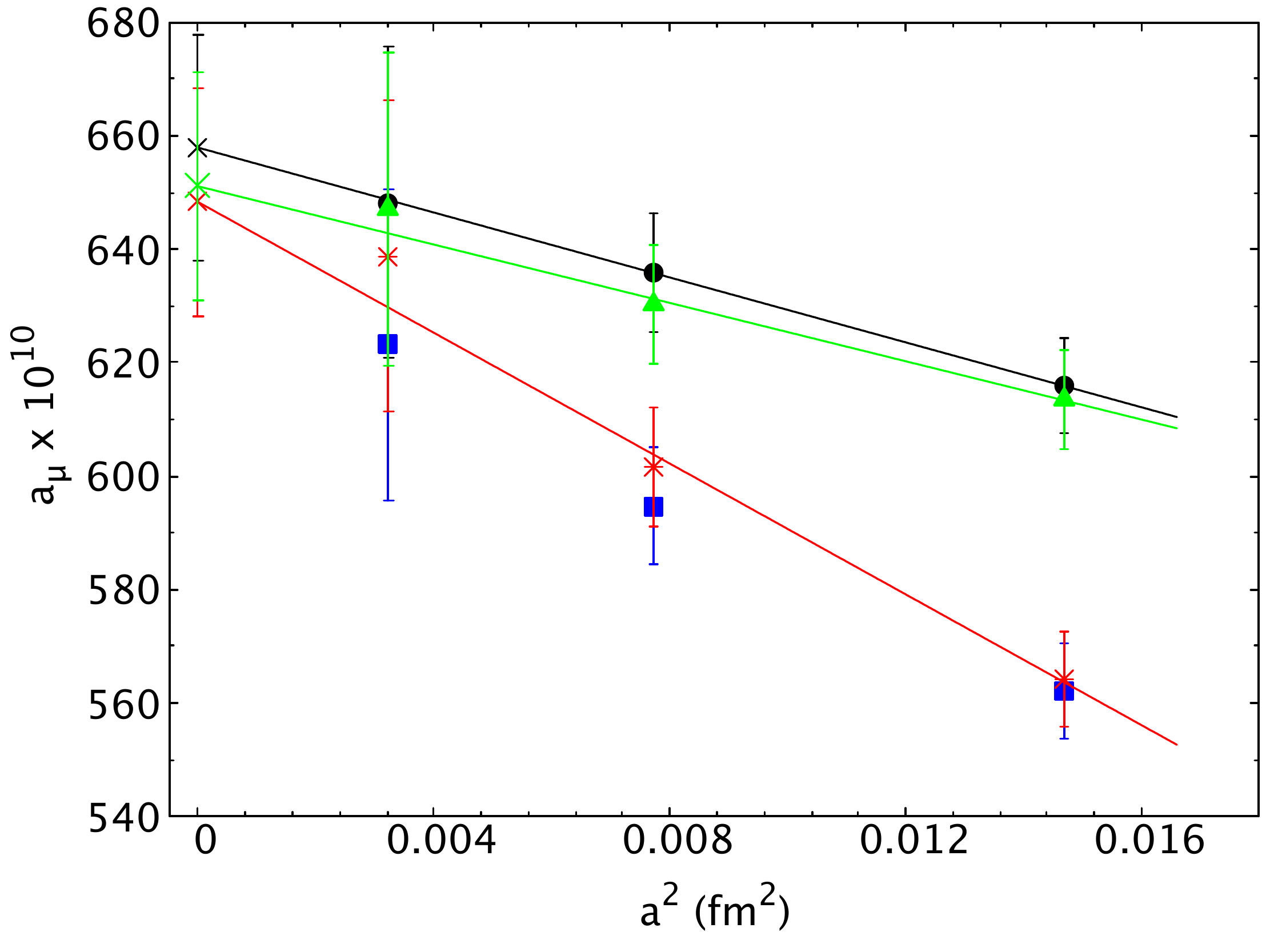}
\caption{Continuum limit of the muon anomaly after correcting the data to infinite volume with NLO staggered chiral perturbation theory (bursts), plus taste corrections (circles), plus pion mass re-tuning (triangles). The uncorrected lattice data (squares) is shown for comparison.}
\label{fig:cont lim}
\end{center}
\end{figure}

In Tab.~\ref{tab:results} we collect results for $a_\mu^{\rm HVP}$ computed on each ensemble, for both $w$ and $\hat w$ weighting functions. Note that scaling violations appear to be smaller for the choice $w$ (see Fig.~\ref{fig:cont lim}).  The muon anomaly for each lattice spacing is shown in Fig.~\ref{fig:cont lim}. Not much is gained from the bounding method for the $48^3$ and $64^3$ ensembles which have small statistical errors already. But on the $96^3$ ensemble there is a clear advantage. The statistical errors in the latter case are larger likely because we have fewer measurements (see Tab.~\ref{tab:ensembles}) and fewer low-modes. For the $96^3$ ensemble, moving the ``averaging window" to the right towards larger times results in larger central values and statistical errors, but with values that are consistent with the one quoted in Tab.~\ref{tab:results}. We chose the range for the central value to avoid the region where the data first fluctuate up, contrary to expectations, while still having significant overlap between the upper and lower bounds. For the other two ensembles, the central value and errors are insensitive to the choice of averaging window.

{To check that the statistical errors are not underestimated, the ensembles were split into halves that were analyzed separately. We expect the errors on the halves to scale roughly like $\sqrt{2}$ times the total error. This scaling was observed except for the $96^3$ ensemble where the first half had an error that was roughly twice the total while the second half was about the same size as the total. In addition the first and second half averages on the $64^3$ ensemble differed by two standard deviations which could signal auto-correlations and an underestimation of the error. We note that the separation between measurements for the $64^3$ ensemble is 12 trajectories while it is 40 and 48 for $48^3$ and $96^3$, respectively. To investigate the $64^3$ ensemble measurements were blocked into averages of one, two, and four consecutive measurements. The errors computed from block sizes two and four were very close to the original analysis. Together, the above suggest the statistical errors quoted here have not been significantly underestimated. }

To take the continuum limit a simple linear in $a^2$ ansatz will be used. But first the data must be corrected for finite volume effects, taste symmetry breaking, and pion mass mistunings (similar corrections were made in Ref.~\cite{Davies:2019efs}). To make the various corrections we employ the following general procedure. The contribution to $a^{\rm HVP}_\mu$ is computed in chiral perturbation theory at NLO. Finite volume corrections are obtained by taking the difference between infinite volume and finite volume results (see Eq.~(\ref{stagDamunum})). Similarly, taste breaking effects are obtained by differences between results computed with the Goldstone pion mass and the average of contributions for each taste pion ($c.f.$, Eq.~(\ref{tastecorrection})). These can be calculated at either finite or infinite volume. Finally, to correct for the mistuning of the pion mass, the difference is computed between the nominal Goldstone mass of 135 MeV and the unitary value measured for each ensemble as given in Tab.~\ref{tab:ensembles}. It turns out the latter correction is only really noticeable for the $64^3$ ensemble (see the fifth column in Tab.~\ref{tab:resultscorrected}), and results in a shift of $-5.71\times 10^{-10}$ from the measured value. This shift is slightly smaller than the one reported in Ref.~\cite{Davies:2019efs} which took the unitary mass to be 128 MeV. Finally, after extrapolating to the continuum and correcting to infinite volume at NLO, we add to the result the average of the NNLO finite volume corrections for each ensemble.

\begin{table}[htp]
\begin{center}
\begin{tabular}{|c|c|c|}
\hline
 $a$ (fm) & total ($w$) & total ($\hat w$) \\
\hline
 0.12121(64) &562.1(8.4) &545.8(8.4)  \\
 0.08787(46) &594.8(10.4)& 584.8(10.4)  \\
 0.05684(30) &623.1(27.5)& 617.8(27.0) \\
 \hline
\end{tabular}
\end{center}
\caption{HVP contributions to the muon anomaly, in units of $10^{-10}$. ``total" refers to the bounding method described in the text, and $w$ ($\hat w$) refers to the use of the weight given by Eq.~(\ref{eq:kernel}) (Eq.~(\ref{eq:what})) in Eq.~(\ref{eq:t-m amu}).}
\label{tab:results}
\end{table}%
\begin{table}[htp]
\begin{center}
\begin{tabular}{|c|c|c|c|c|}
\hline
 $a$ (fm) & lattice value & FV corr. & FV + taste corr. & FV+taste+$m_\pi$ corr. \\
\hline
 0.12121(64) &562.1(8.4) & 564.2(8.4)  &615.8(8.4)& 613.6(8.4)\\
 0.08787(46) &594.8(10.4)& 601.7(10.4)  &  635.9(10.4)& 630.2(10.4)\\
 0.05684(30) &623.1(27.5)& 638.7(27.5) & 648.2(27.5) & 647.1(27.5)\\
 \hline
0 &  & 648.3(20.0) & 657.9(20.0)& 651.1(20.1)\\
\hline
\end{tabular}
\end{center}
\caption{HVP contributions to the muon anomaly, in units of $10^{-10}$, including corrections computed in chiral perturbation theory. The second column repeats the second column of
Table~\ref{tab:results}, the third column includes the finite-volume corrections of Eq.~(\ref{stagDamunum}), while
the fourth column also includes the infinite-volume taste corrections of Eq.~(\ref{tastecorrection}).
The fifth column adjusts the values shown in the fourth column to a common pion mass of 135~MeV using NLO ChPT, as described in the text.
Continuum extrapolated values of each column are shown in the last row. The weighting function $w$ has been used throughout.}
\label{tab:resultscorrected}
\end{table}%

In Tab.~\ref{tab:resultscorrected}, values of $a_\mu^{\rm HVP}$, including finite volume and finite volume plus taste corrections for each ensemble, are given in the third and fourth columns, respectively. They are also displayed in Fig.~\ref{fig:cont lim}. Values in the continuum should agree, so the difference is a measure of the systematic error associated with the continuum extrapolation, which we take as one-half of the difference, which is equal to $4.8\times10^{-10}$. The fifth column gives $a_\mu^{\rm HVP}$ after NLO corrections for finite volume, taste symmetry breaking, and pion mass re-tuning, which we take as the NLO-corrected central value.  
Applying the averaged NNLO finite volume correction of $8\times 10^{-10}$ from Eq.~(\ref{DamuNNLOnumtotal}) with a ChPT error of $4\times 10^{-10}$ to this result then yields
\begin{equation}
\label{eq:final value}
(659\pm 20\pm 5\pm 5\pm 4)\times 10^{-10}=659(22)\times 10^{-10}.
\end{equation}
The first, dominant, error is statistical, while the rest are systematic error estimates (in order of size): continuum extrapolation, scale setting,\footnote{For the values of $a$ given in Tab.~\ref{tab:ensembles}, we simply adopt the scale setting error given in Tab. IV of \cite{Davies:2019efs}.} and higher orders in ChPT. The second equation gives the error by adding the individual ones in quadrature.

\begin{figure}[htbp]
\begin{center}
\includegraphics[width=0.7\textwidth]{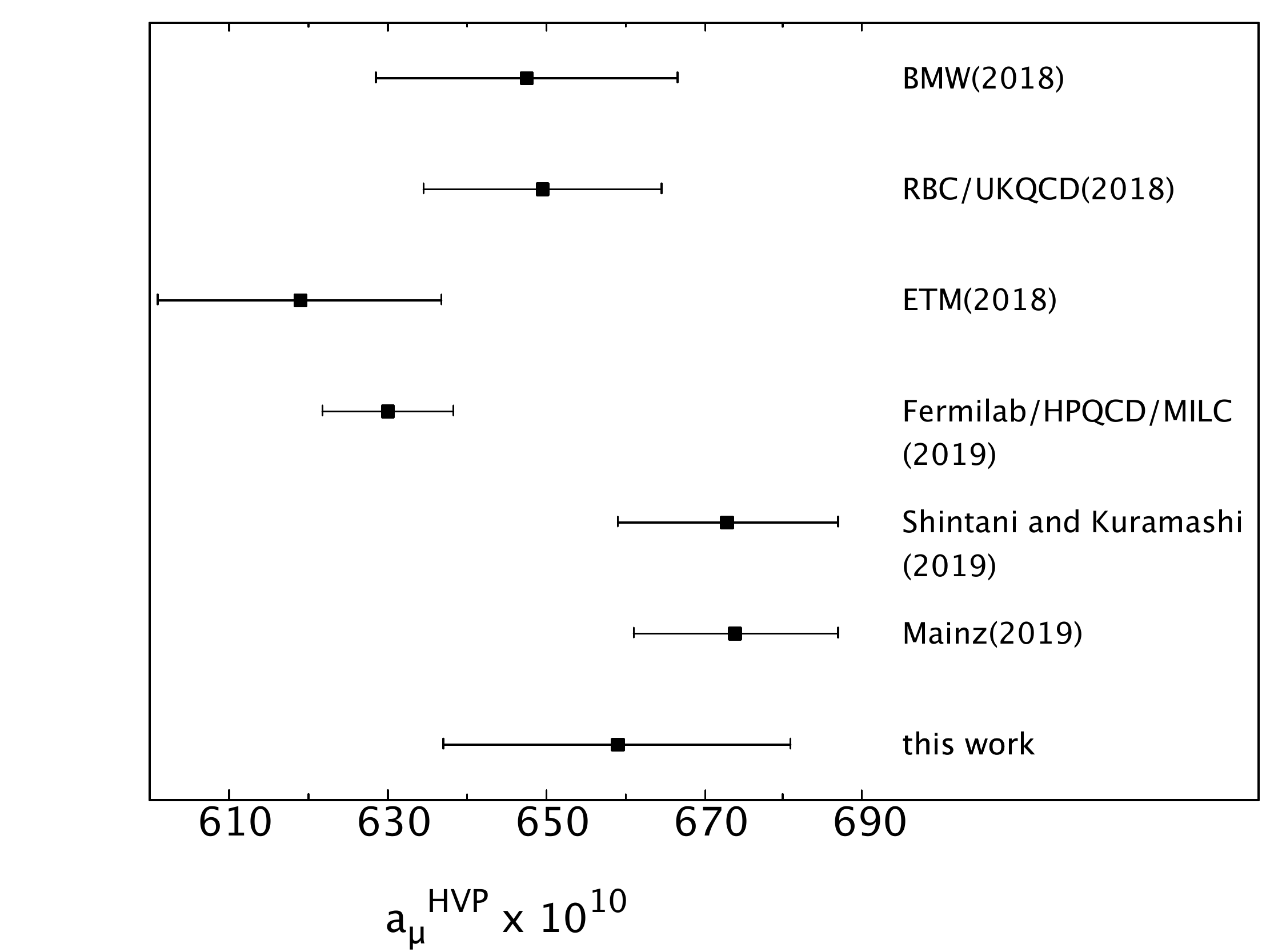}
\caption{Contributions to the muon anomaly from the connected light quark vacuum polarization from recent publications \cite{Borsanyi:2017zdw} (BMW), \cite{Blum:2018mom} (RBC/UKQCD), \cite{Giusti:2018mdh} (ETM), \cite{Davies:2019efs} (Fermilab/HPQCD/MILC), \cite{Shintani:2019wai} (Shintani and Kuramashi), \cite{Gerardin:2019rua} (Mainz).}
\label{fig:amu lit}
\end{center}
\end{figure}

The FNAL/MILC/HPQCD collaborations recently produced an update of their computation of the HVP contribution~\cite{Davies:2019efs}, using the same physical mass HISQ ensembles as those employed here (plus two additional ones with $a\approx 0.15$ fm), so it is particularly interesting to compare our results with that work. Those authors use different methods, including moments of local-local current correlation functions and Pad\'e approximants~\cite{Aubin:2012me,Chakraborty:2016mwy}. They do not use LMA, instead relying on brute-force computations on 1000's of configurations to control statistical errors. Because our computations are so different, consistency is a significant test of these lattice computations. The values of (uncorrected) light quark connected contribution are given in Tab. III of Ref.~\cite{Davies:2019efs} for the three ensembles used in this work. They find\footnote{The errors given here are statistical only (private communication with the authors). In Tab. III of Ref.~\cite{Davies:2019efs} the errors are statistical and systematic, combined in quadrature.} 580(7), 605(7), and 608(14) in units of $10^{-10}$ compared to the values in the second column of Tab.~\ref{tab:results}, 562(8), 595(10), and 623(28). All of the errors just quoted are statistical only, and comparable, except for the $96^3$ ensemble. Since the lattice spacing errors in the valence quark sector are different between the two calculations, {the above values need not agree precisely except in the continuum and infinite volume limits.} The value quoted in Eq.~(3.2) of Ref.~\cite{Davies:2019efs} is 630.1(8.3) which is consistent, but somewhat smaller than, the value given in Eq.~(\ref{eq:final value}).
{The authors of Ref.~\cite{Davies:2019efs} also use a prior constraint on the coefficient of the $a^2$ term which reduces the uncertainty on the continuum limit extrapolation.
At closer inspection the results on each ensemble are not so different either. The points at 0.09 and 0.12 fm show similar behavior, and 
it could be informative to obtain the point at 0.15 fm using our method to better compare the overall $a^2$ dependence. The 0.06 fm points also agree well within (larger) statistical errors. Finally, a significant part of the difference between the values comes from the corrections beyond NLO ChPT: ours is $+8\times10^{-10}$, coming from NNLO ChPT, while their model estimate varies from $-4\times10^{-10}$ to $-10\times10^{-10}$, depending on the ensemble.
Our result is consistent within errors with other recent computations, as seen in Fig.~\ref{fig:amu lit}. However there is still a relatively large spread, with the values on the low and high ends being incompatible with each other.
}

{To explore a more precise comparison with results from other groups, we adopt the window method of Ref.~\cite{Blum:2018mom}: }
\begin{eqnarray}
a_\mu^{W} &=& 2\sum_{t=0}^{T/2} C(t) w(t) (\Theta(t,t_0,\Delta)-\Theta(t,t_1,\Delta))\\
\Theta(t,t',\Delta) &=& \frac{1}{2} (1+\tanh((t-t')/\Delta))
\end{eqnarray}
where $t_1-t_0$ is the size of the window and $\Delta$ is a suitably chosen width that smears out the window at either edge. We choose windows to avoid both lattice artifacts at short distance and large statistical errors at long distance. Results for several windows and both weighting functions are tabulated in Tab.~\ref{tab:window results}. {We note that the window method can also be used to combine lattice and dispersive results to obtain a result that is more precise than either alone as was shown in Ref.~\cite{Blum:2018mom}, though we do not pursue this here.}

\begin{table}[htp]
\begin{center}
\begin{tabular}{|c|c|c|c|c|c|c|}
\hline
 $a$ (fm) & window 1& window 2 & window 3 & window 1($\hat w$)& window 2($\hat w$) & window 3 ($\hat w$)\\
\hline
 0.12121(64)  & 201.07(56) & 186.43(51)&308.32(94) & 194.12(55) &179.32(49) & 300.20(93)\\
 0.08787(46)  & 205.95(66)& 191.89(69) &319.16(1.44) &202.22(65) & 187.95(68) & 314.79(1.42)\\
 0.05684(30) &  207.13(92)& 193.91(1.02)&324.37(2.40) &205.55(91) & 192.18(1.02)&322.52(2.39) \\
 \hline
0 & 209.78(96) & 196.82(1.03) & 329.99(2.25)&209.69(95) & 196.52(1.02)& 329.85(2.24)\\
\hline
\end{tabular}
\end{center}
\caption{HVP contributions to the muon anomaly, in units of $10^{-10}$, from the window method with windows 1, 2, and 3, $(t_0,t_1,\Delta)=(0.4,1.0,0.15)$, (0.4,1.0,0.3), and (0.4,1.3,0.15), respectively. $\hat w$ refers to the weighting function (\ref{eq:what}) in Eq.~(\ref{eq:t-m amu}).}
\label{tab:window results}
\end{table}%
\begin{figure}[htbp]
\begin{center}
\includegraphics[width=0.75\textwidth]{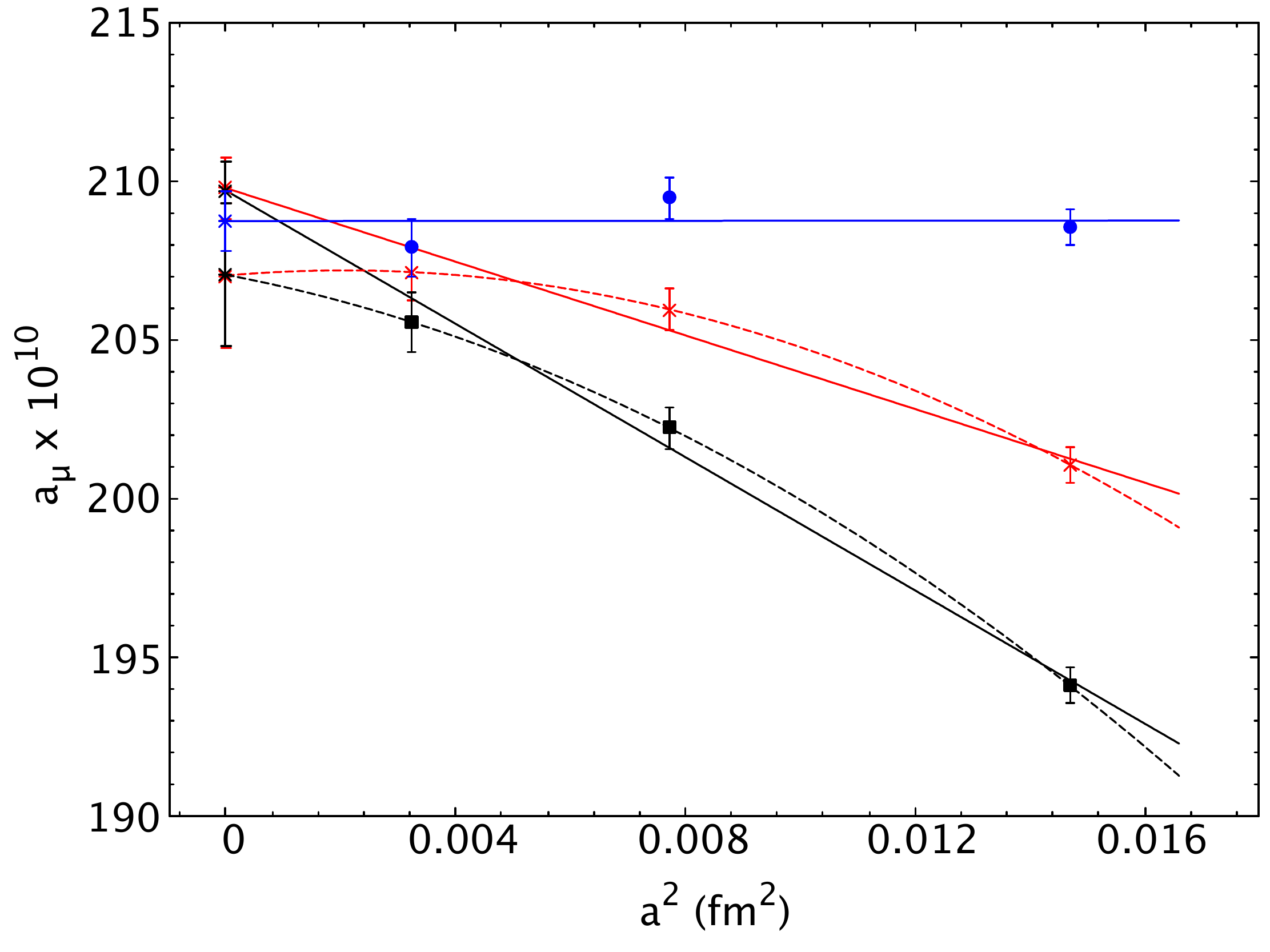}
\caption{Continuum limit combined with the window method for lattice data without finite volume corrections. $t_0=0.4$ fm, $t_1=1$ fm, $\Delta=0.15$. 
Squares (crosses) correspond to uncorrected data points with weighting function $\hat w$ ($w$); filled circles are taste-breaking corrected to NLO from $w$ data points.  Solid curves show linear fits in $a^2$; all three agree very well in the continuum limit. Dashed curves  denote a fully constrained parametrization (no degrees of freedom) using both $a^2$ and $a^4$ terms. 
}
\label{fig:window1}
\end{center}
\end{figure}

In Fig.~\ref{fig:window1} several continuum limits are shown for the window with $t_0=0.4$, $t_1=1$, and $\Delta=0.15$ fm. For this window the statistical errors for each ensemble are very small, so it allows a precise regime to explore and understand discretization effects. Here we also ignore mass re-tunings and finite volume effects because they have a negligible effect, with the two-pion state dominating only at long distance (an explicit check reveals this assertion to be true). However, we do investigate taste-breaking effects since these are significant. The lower two curves in Fig.~\ref{fig:window1} correspond to uncorrected data points and weighting functions $w$ and $\hat w$. At non-zero lattice spacing there is a noticeable effect, but the continuum limits are the same (see the last row in Tab.~\ref{tab:window results}). Including the taste-breaking corrections shifts the data further, essentially making the curve flat, but the continuum limit is barely affected. We also show totally constrained ``fits," including an $a^4$ term, which lower the continuum limit slightly while significantly increasing the statistical error. 
Linear extrapolations using only the two finer ensembles give very similar results. {The various values, which are very different at non-zero lattice spacing, and different extrapolations give consistent results in the continuum limit, with small differences that are well within statistical errors.}

\begin{figure}[htbp]
\begin{center}
\includegraphics[width=0.7\textwidth]{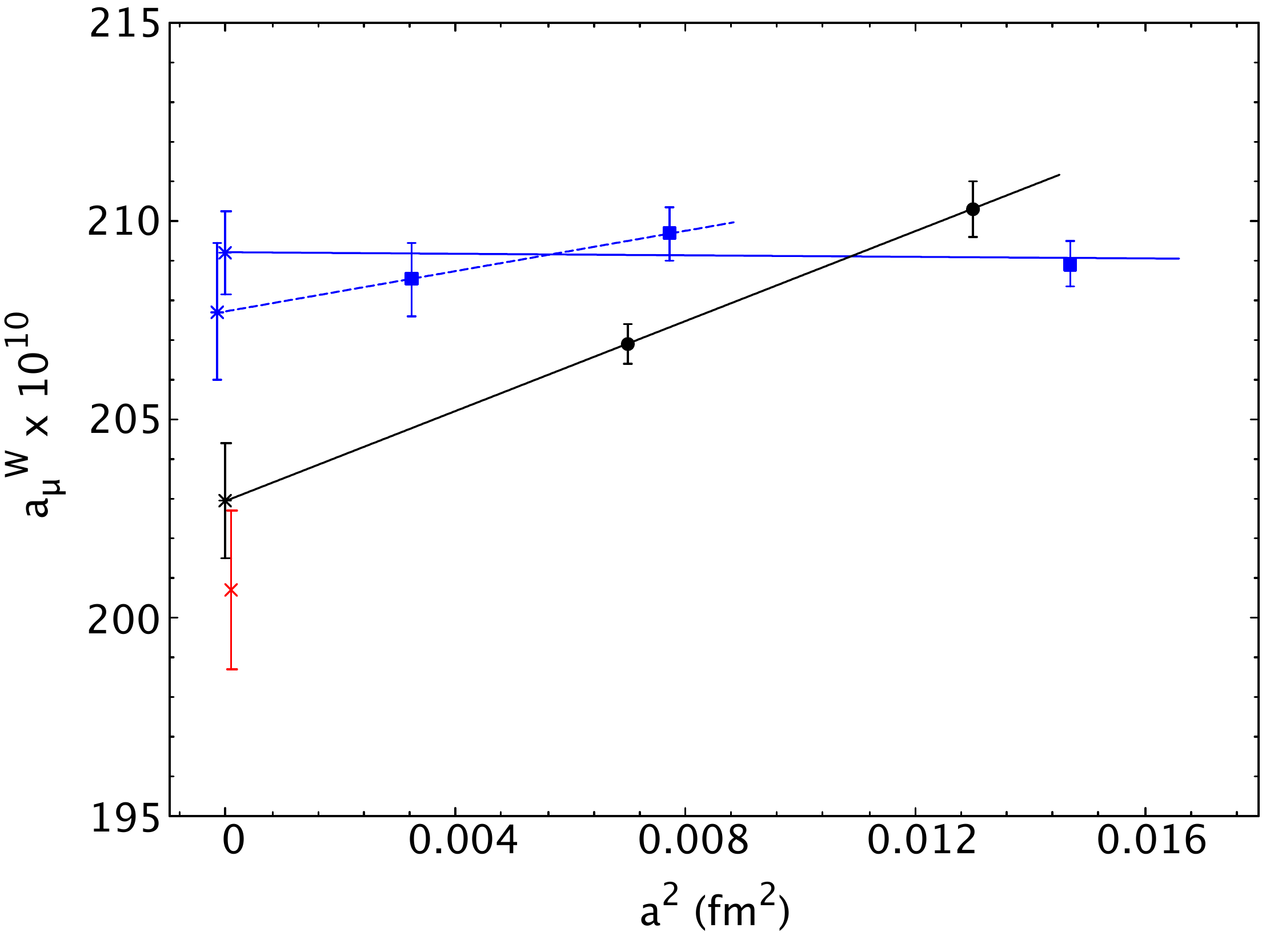}
\includegraphics[width=0.7\textwidth]{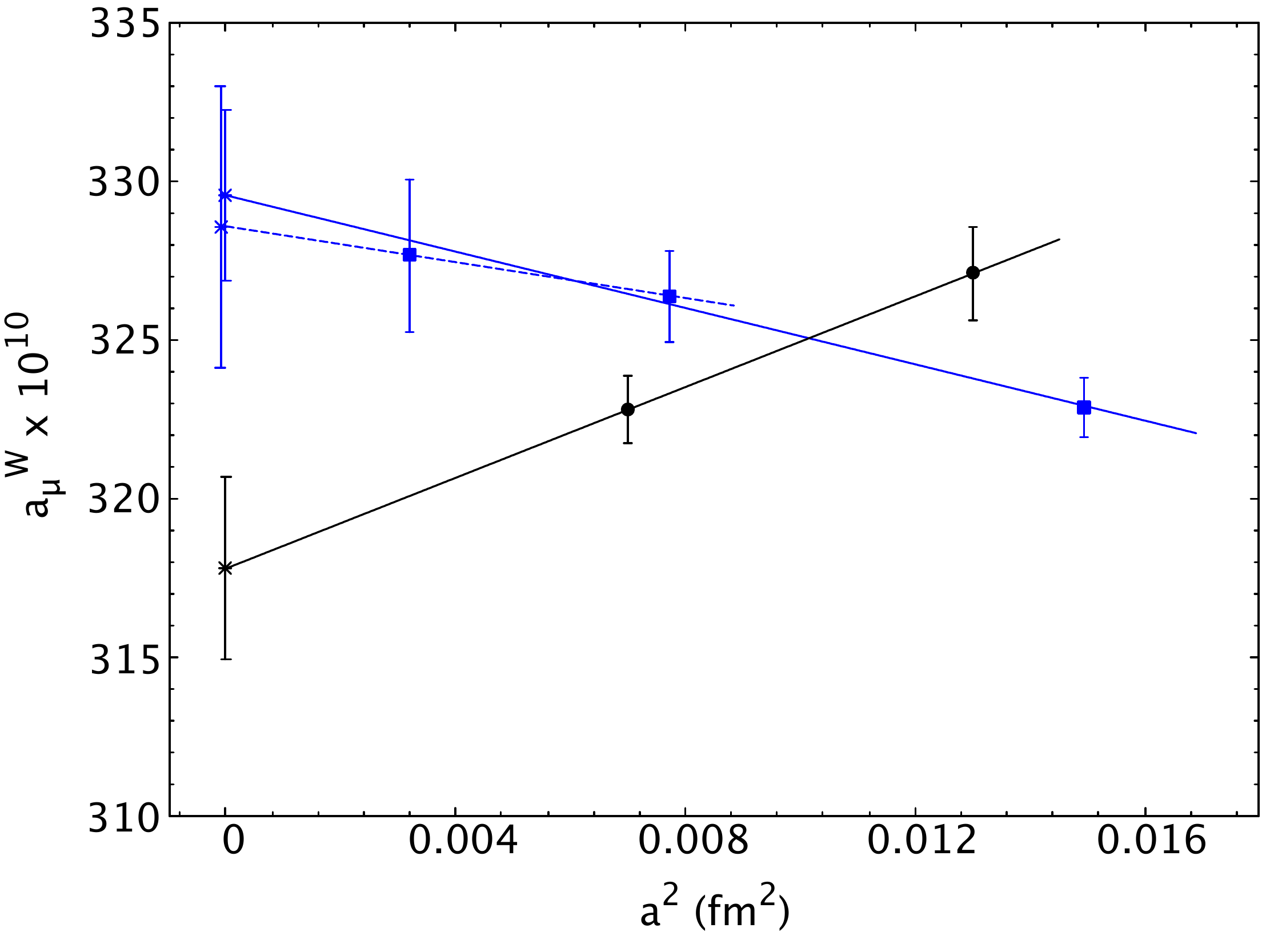}
\caption{Continuum limit combined with the window method for DWF~\cite{Blum:2018mom}{, using the weight $\hat{w}$} (circles) and HISQ{, using the weight $w$} (squares). $\Delta=0.15$, $t_0=0.4$ fm, $t_1=1$ fm (upper panel) and 1.3 fm (lower panel). The R-ratio result (cross, using data from Ref.~\cite{Keshavarzi:2018mgv} by C. Lehner) is also shown in the upper panel. Finite volume (DWF and HISQ) and taste breaking (HISQ) corrections have been included to NLO in ChPT. Lattice spacing uncertainties, added in quadrature with statistical errors, are also included.}
\label{fig:window}
\end{center}
\end{figure}

Figure~\ref{fig:window} displays results for two representative windows along with values from the recent RBC/UKQCD computation using domain wall fermions (DWF)~\cite{Blum:2018mom}\footnote{Here we compare results that have been corrected to NLO in ChPT since that is what is published in Ref.~\cite{Blum:2018mom}.}. The results should agree in the continuum limit. We also show the corresponding dispersive/$e^+e^-$ value, using the R-ratio compilation of Ref.~\cite{Keshavarzi:2018mgv}.
The HISQ results lie above the DWF and dispersive ones. The differences in central values correspond to roughly 1-2 percent of the total HVP contribution to $a_\mu$, depending on the window and the fit. Leaving out the largest lattice spacing point (for HISQ) tends to give a somewhat lower value with larger statistical errors. Given the uncertainties (2-4 standard deviations) it is difficult to conclude if there is a significant discrepancy, though the spread seems uncomfortably large. The statistical error on the $a=0$ HISQ point is smaller than for DWF, but the uncertainty due to the lattice spacing is larger. When adding them in quadrature, the total errors for HISQ are smaller for window 1 and about the same for window 3. It is interesting to note that the HISQ and DWF lattice spacing errors are comparable before taste symmetry breaking corrections, and that after including corrections the HISQ points are remarkably flat, especially for window 1. In Fig.~\ref{fig:window} finite volume errors have been included to NLO, but are very small in both windows and shift both DWF and HISQ curves up by roughly the same amount. The absence of charm sea quarks in the DWF result is estimated from perturbation theory to be very small~\cite{Blum:2018mom}.

{To check if the errors above were underestimated for the windows, we performed the first half -- second half analysis as before. The situation turns out to be similar to the case for the total, except the values for the $64^3$ ensemble which are now closer to three sigma away from each other. If we inflate the errors by 50\% on all points, then the difference is again below 2 standard deviations, the error on the $a=0$ value also grows by 50\%, and the almost 4 sigma discrepancy found above goes down to about 3.5 sigma. Thus our conclusions remain unchanged.}

{One can see a tension between the HISQ and DWF results in the continuum limit from this comparison. Whether or not the difference will survive after further investigation is unclear at this point. A third, smaller, lattice spacing ensemble is being generated by the RBC/UKQCD collaborations~\cite{rbcukqcd}, and we plan to add statistics and a fourth lattice spacing in the future, both of which should help resolve the issue.} It would be helpful if other groups also applied the window method to their existing data.

A final check included for completeness comes from moments of the correlation function~\cite{Chakraborty:2016mwy},
\begin{eqnarray}
\Pi^{ll}_n &=& (Q_u^2+Q_d^2)(-1)^{n+1}~2\sum_{t=0}^{T/2} \frac{t^{2n+2}}{(2n+2)!}C(t).
\end{eqnarray}
For the first moment we find 0.0797(27), 0.0841(39), and 0.069(39) for the three different ensembles, coarsest to finest, respectively. A simple linear extrapolation in $a^2$ yields $\Pi^{ll}_1=0.0884(86)$ which is consistent with the values in Refs.~\cite{Davies:2019efs,Blum:2018mom}.

\section{Conclusion}
\label{sec:conclusion}

We have presented a lattice QCD computation of the light quark HVP contribution to the muon anomaly with 2+1+1 flavors of HISQ fermions. Three ensembles at the physical point, generated by the MILC Collaboration, were used to take the continuum limit at fixed volume ($L\approx 5.5$ fm), and the results are broadly consistent with those in the literature. Using the window method, a precise comparison yields values that are a bit higher than the dispersive result and a recent one using DWF. Given the statistical and systematic errors it is not clear that a real discrepancy exists: a decisive determination requires additional computations. 

Overall the statistical errors in this study are at the larger end of the range from recent studies~\cite{DellaMorte:2017dyu,Blum:2002ii,Borsanyi:2017zdw,Giusti:2018mdh,Blum:2018mom,Davies:2019efs}. This is primarily due to the fewer number of low modes and measurements on the largest lattice used in our study. 
Nevertheless the error reduction techniques used here are demonstrably powerful. Future computations with more measurements, and in particular, that use more low modes, can have an impact.

We have also presented a calculation in chiral perturbation theory, in Euclidean space, through NNLO of the finite volume corrections to the HVP contribution to the muon $g-2$. The NNLO correction is large ($\sim$ 1\%) for physical pion mass and the lattice sizes used in current calculations, so it must be included for a precise comparison to experiment.

The computations presented here are important for the test of the Standard Model against the ongoing experiment at Fermilab and an upcoming one at J-PARC.

\section{Acknowledgments}
{We thank Ruth Van de Water for discussions on the continuum limit fit in Ref.~\cite{Davies:2019efs}, Christoph Lehner for discussions and providing DWF and R-ratio data on the window method, and the MILC collaboration for the use of their gauge configurations.}
This work was partially supported by the US DOE. Computational resources were provided by the USQCD Collaboration. T.B. and C.T., and M.G. were supported in part by the U.S. Department of Energy under Awards No. DE-FG02-92ER40716 and No. DE-FG03-92ER40711, respectively. C.J. was supported in part by the U.S. Department of Energy contract DE-SC0012704.
S.P. is supported by CICYTFEDER-FPA2017-86989-P and by Grant No. 2017 SGR 1069. The software used for this work includes  \href{https://github.com/RBC-UKQCD/CPS}{CPS} and
\href{https://github.com/paboyle/Grid}{Grid}.

\section*{Appendix:  NNLO finite-volume correction}
At NNLO, using the resummation formula~(\ref{Presum}), $C(t)$ of Eq.~(\ref{CtNNLOdimreg})
can be rewritten as
\begin{eqnarray}
\label{Ctdimregrewrite}
C(t)
&=&-\frac{10}{9}\Biggl(\frac{1}{3}\,\int \frac{d^dp}{(2\pi)^d}\frac{{\vec p}^2}{E_p^2}\,e^{-2E_p t}\Biggl[1-\frac{8({\vec p}^2+m_{\pi}^2)}{F^2}\,\ell_6
-\frac{1}{F^2}\int \frac{d^dk}{(2\pi)^d}\frac{1}{E_k}\Biggr]\\
&&+\frac{1}{6dF^2}\int \frac{d^dp}{(2\pi)^d}\int \frac{d^dk}{(2\pi)^d}
\frac{{\vec p}^2{\vec k}^2}{E_p^2E^2_k}\frac{E_k e^{-2E_p t}-E_p e^{-2E_k t}}{{\vec k}^2-{\vec p}^2}
\nonumber\\
&&+\frac{1}{3}\,\int \frac{d^dp}{(2\pi)^d}\sum_{{\vec n\ne 0}}e^{i{\vec n}\cdot{\vec p}L}\frac{{\vec p}^2}{E_p^2}\,e^{-2E_p t}\Biggl[1-\frac{8({\vec p}^2+m_\pi^2)}{F^2}\,\ell_6-\frac{1}{F^2}\int \frac{d^dk}{(2\pi)^d}\frac{1}{E_k}\Biggr]\nonumber\\
&&-\frac{1}{3F^2}\,\int \frac{d^dp}{(2\pi)^d}\frac{{\vec p}^2}{E_p^2}\,e^{-2E_p t}\Biggl[\int \frac{d^dk}{(2\pi)^d}\sum_{\vec m\ne 0}\frac{e^{i{\vec m}\cdot{\vec k}L}}{E_k}\Biggr]\nonumber\\
&&+\frac{1}{3dF^2}\int \frac{d^dp}{(2\pi)^d}\sum_{\vec n\ne 0}e^{i{\vec n}\cdot{\vec p}L}\int \frac{d^dk}{(2\pi)^d}
\frac{{\vec p}^2{\vec k}^2}{E_p^2E^2_k}\frac{E_k e^{-2E_p t}-E_p e^{-2E_k t}}{{\vec k}^2-{\vec p}^2}
\nonumber\\
&&-\frac{1}{3F^2}\,\int \frac{d^dp}{(2\pi)^d}\sum_{{\vec n\ne 0}}e^{i{\vec n}\cdot{\vec p}L}\frac{{\vec p}^2}{E_p^2}\,e^{-2E_p t}\Biggl[\int \frac{d^dk}{(2\pi)^d}\sum_{\vec m\ne 0}\frac{e^{i{\vec m}\cdot{\vec k}L}}{E_k}\Biggr]\nonumber\\
&&+\frac{1}{6dF^2}\int \frac{d^dp}{(2\pi)^d}\int \frac{d^dk}{(2\pi)^d}\sum_{{\vec n\ne 0},{\vec m\ne 0}}
e^{i{\vec n}\cdot{\vec p}L+i{\vec m}\cdot{\vec k}L}
\frac{{\vec p}^2{\vec k}^2}{E_p^2E^2_k}\frac{E_k e^{-2E_p t}-E_p e^{-2E_k t}}{{\vec k}^2-{\vec p}^2}\Biggr)\ .
\nonumber
\end{eqnarray}
We note that, despite the appearance of ${\vec k}^2-{\vec p}^2$ in the denominator in various
places, this is always accompanied by a numerator that vanishes at ${\vec k}^2={\vec p}^2$, and
all functions we integrate over ${\vec k}$ and ${\vec p}$ are continuous.   An implication is that if
we (as we will do below) break up some of the terms containing the factor $1/({\vec k}^2-{\vec p}^2)$,
any contributions from the apparent pole at ${\vec k}^2={\vec p}^2$ should be dropped.   We will
always regulate such poles such that they do not contribute to the integrals.

The first two lines give the infinite-volume result, while the remaining lines represent finite-volume
corrections.   These finite-volume corrections can be rearranged as
\begin{eqnarray}
\label{FVrewrite}
\Delta C(t)&=&
-\frac{10}{9}\Biggl(\frac{1}{3}\int \frac{d^3p}{(2\pi)^3}\sum_{{\vec n\ne 0}}e^{i{\vec n}\cdot{\vec p}L}\frac{{\vec p}^2}{E_p^2}\,e^{-2E_p t}\Biggl[1-\frac{m_\pi^2}{36\pi^2 F^2}+\frac{5{\vec p}^2}{36\pi^2 F^2}
+\frac{{\vec p}^2+m_\pi^2}{12\pi^2 F^2}\,\overline{\ell}_6\\
&&\hspace{5.2cm}-\frac{{\vec p}^2}{6\pi^2 F^2}\sqrt{\frac{{\vec p}^2}{{\vec p}^2+m_\pi^2}}\log\left(\sqrt{\frac{{\vec p}^2}{m_\pi^2}}+\sqrt{\frac{{\vec p}^2}{m_\pi^2}+1}\right)\Biggr]\nonumber\\
&&-\frac{1}{3F^2}\,\int \frac{d^dp}{(2\pi)^d}\sum_{{\vec n\ne 0}}e^{i{\vec n}\cdot{\vec p}L}\frac{{\vec p}^2}{E_p^2}\,e^{-2E_p t}\Biggl[\int \frac{d^dk}{(2\pi)^d}\sum_{\vec m\ne 0}\frac{e^{i{\vec m}\cdot{\vec k}L}}{E_k}\Biggr]\nonumber\\
&&-\frac{1}{3F^2}\int \frac{d^dp}{(2\pi)^d}\frac{{\vec p}^2}{E_p^2}\,e^{-2E_p t}\Biggl[\int \frac{d^dk}{(2\pi)^d}\sum_{\vec n\ne 0}\frac{e^{i{\vec n}\cdot{\vec k}L}}{E_k}\left(1-\frac{1}{d}
\frac{{\vec k}^2}{{\vec k}^2-{\vec p}^2}\right)\Biggr]\nonumber\\
&&+\frac{1}{6dF^2}\int \frac{d^dp}{(2\pi)^d}\int \frac{d^dk}{(2\pi)^d}\sum_{{\vec n\ne 0},{\vec m\ne 0}}
e^{i{\vec n}\cdot{\vec p}L+i{\vec m}\cdot{\vec k}L}
\frac{{\vec p}^2{\vec k}^2}{E_p^2E^2_k}\frac{E_k e^{-2E_p t}-E_p e^{-2E_k t}}{{\vec k}^2-{\vec p}^2}\Biggr)\ ,
\nonumber
\end{eqnarray}
in which the renormalization-group invariant $\overline{\ell}_6$ is defined by
\begin{equation}
\label{bell6}
\ell_6^r(\mu)=-\frac{1}{96\pi^2}\left(\overline{\ell}_6+\log\frac{m_\pi^2}{\mu^2}\right)\ ,
\end{equation}
and where the limit $d\to 3$ has already been taken in the first term.
The first term in Eq.~(\ref{FVrewrite}) collects the terms containing the factors $e^{-2E_p t}$ on the
third and fifth lines of Eq.~(\ref{Ctdimregrewrite}), the second term (third line) collects the fourth
line and the remaining part of the fifth line (with the interchange ${\vec p}\leftrightarrow{\vec k}$),
while the last two lines are copied from the sixth and seventh lines of Eq.~(\ref{Ctdimregrewrite}).

The first term (first two lines) of Eq.~(\ref{FVrewrite}) can be dealt with in the same way as the
NLO contribution; all one needs to do is to insert the expression between square brackets inside
the integral over $p$ in Eq.~(\ref{amuNLO}).   Numerically, using
\begin{eqnarray}
\label{Fell6}
F=F_\pi&=&92.21\ \mbox{MeV}\ ,\\
\overline{\ell}_6&=&16(1)\qquad (\mbox{Ref. \cite{BCT}})\ ,\nonumber
\end{eqnarray}
we find that this shifts the values
we found in Eq.~(\ref{DamuNLOnum}) by
\begin{equation}
\label{DamuNNLOnum1}
\Delta a_\mu^{\rm HVP,\,NNLO,\,1}=\left\{\begin{array}{c} 8.89\times 10^{-10}\ ,\quad L/a=96\\
8.77\times 10^{-10}\ ,\quad L/a=64\\
7.22\times 10^{-10}\ ,\quad L/a=48
\end{array}\right.\ .
\end{equation}
For the third and fourth lines in Eq.~(\ref{FVrewrite}), we need the integral
\begin{equation}
\label{bessel}
\frac{1}{F^2}\int \frac{d^dk}{(2\pi)^d}\sum_{\vec n\ne 0}\frac{e^{i{\vec n}\cdot{\vec k}L}}{E_k}=\frac{1}{F^2}\frac{\pi^{d/2}}{(2\pi)^d\Gamma(d/2)}\sum_{\vec n\ne 0}\frac{1}{inL}\int_{-\infty}^\infty (k^2)^{\epsilon/2}dk\,\frac{k}{\sqrt{k^2+m_\pi^2}}\,e^{inkL}\ ,
\end{equation}
which converges for $d<3$.   We use Cauchy's theorem to rewrite the $k$ integral as an integral along the discontinuity
of the square root across the cut which we choose along the positive imaginary axis starting at $+im_\pi$.\footnote{There is also a branch cut starting at $-im_\pi$ which we can take along the negative imaginary axis.   The branch point at $k=0$ does not contribute in the limit $\epsilon\to 0$.} The result is finite in the limit $d\to 3$, 
and Eq.~(\ref{bessel}) then becomes equal to
\begin{equation}
\label{bessel2}
\frac{1}{2\pi^2F^2}\sum_{\vec n\ne 0}\frac{1}{nL}\int_{m_\pi}^\infty dy\frac{y}{\sqrt{y^2-m_\pi^2}}\,e^{-ynL}
=-\frac{m_\pi^2}{2\pi^2F^2}\sum_{n^2=1}^\infty\frac{Z_{00}(0,n^2)}{nm_\pi L}K_1(nm_\pi L)\ .
\end{equation}
The numerical value of this expression is equal to $0.00399$, $0.00375$ and $0.00282$ for
the $96^3$, $64^3$ and $48^3$ ensembles, respectively.   From these numbers, and using the values of Eq.~(\ref{DamuNLOnum}), we find for the contribution from the third line of Eq.~(\ref{FVrewrite}) the values
\begin{equation}
\label{DamuNNLOnum2}
\Delta a_\mu^{\rm HVP,\,NNLO,\,2}=\left\{\begin{array}{c} -0.08\times 10^{-10}\ ,\quad L/a=96\\
-0.08\times 10^{-10}\ ,\quad L/a=64\\
-0.05\times 10^{-10}\ ,\quad L/a=48
\end{array}\right.\ .
\end{equation}

The other integral over ${\vec k}$ on the fourth line of Eq.~(\ref{FVrewrite}) is, writing $k^2={\vec k}^2$
and $p^2={\vec p}^2$,
equal to
\begin{eqnarray}
\label{2ndintegral}
B(p^2)&\equiv&\lim_{\eta\to 0}\lim_{d\to 3}\frac{1}{F^2}\int \frac{d^dk}{(2\pi)^d}\sum_{\vec n\ne 0}\frac{e^{i{\vec n}\cdot{\vec k}L}}{E_k}
\frac{k^2}{   k^2-p^2+2i\eta k}\\
&=&\lim_{\eta\to 0}\lim_{d\to 3}\frac{1}{F^2}\frac{\pi^{d/2}}{(2\pi)^d\Gamma(d/2)}\sum_{\vec n\ne 0}\frac{1}{inL}\int_{-\infty}^\infty (k^2)^{\epsilon/2}dk\,\frac{k}{\sqrt{k^2+m_\pi^2}}\,e^{inkL}\frac{k^2}{k^2-p^2+2i\eta k}
\nonumber\\
&=&-\frac{m_\pi^2}{2\pi^2F^2}\sum_{n^2=1}^\infty\frac{Z_{00}(0,n^2)}{nm_\pi L}\int_1^\infty dy\frac{y}{\sqrt{y^2-1}}\frac{y^2}{y^2+\frac{{\vec p}^2}{m_\pi^2}}\,e^{-ynm_\pi L}\ .\nonumber
\end{eqnarray}
Here we again closed the contour in the upper half $k$ plane, and regulated the poles at $k=
\pm p-i\eta$
such that they are located in the lower half $k$ plane, and thus do not contribute; {\it cf.} the explanation
below Eq.~(\ref{Ctdimregrewrite}).

Using Eqs.~(\ref{bessel2}) and~(\ref{2ndintegral}) to numerically carry out the integral over ${\vec p}$ on the
fourth line of Eq.~(\ref{FVrewrite}), we find the corrections
\begin{equation}
\label{DamuNNLOnum3}
\Delta a_\mu^{\rm HVP,\,NNLO,\,3}=\left\{\begin{array}{c} 0.30\times 10^{-10}\ ,\quad L/a=96\\
0.30\times 10^{-10}\ ,\quad L/a=64\\
0.22\times 10^{-10}\ ,\quad L/a=48
\end{array}\right.\ .
\end{equation}
The final term in Eq.~(\ref{FVrewrite}) can be brought into a simpler form by carrying out the
angular integrals, leading to a contribution to $\Delta a_\mu^{\rm HVP}$ of the form
\begin{eqnarray}
\label{FVNNLO3}
\Delta a_\mu^{\rm HVP,\,NNLO,\,4}&=&\frac{10}{9}
\frac{\alpha^2}{24dF^2}
\left(\frac{2\pi^{d/2}}{\Gamma(d/2)(2\pi)^d}\right)^2\sum_{n^2=1}^\infty\sum_{m^2=1}^\infty\frac{Z_{00}(0,n^2)Z_{00}(0,m^2)}{nmL^2}\\
&&\hspace{0cm}\times
\int_{-\infty}^\infty p^{d-3}dp\int_{-\infty}^\infty k^{d-3}dk\,e^{inpL+imkL}\frac{p^3k^3}{E_pE_k}\frac{F(p^2)/E_p-F(k^2)/E_k}{k^2-p^2}\ .
\nonumber
\end{eqnarray}
Interchanging $p$ and $k$ in the
integral with $F(k^2)/E_k$ in the numerator, we obtain
\begin{eqnarray}
\label{FVNNLO3p}
\Delta a_\mu^{\rm HVP,\,NNLO,\,4}&=&\frac{10}{9}
\frac{\alpha^2}{12dF^2}
\left(\frac{2\pi^{d/2}}{\Gamma(d/2)(2\pi)^d}\right)^2\sum_{n^2=1}^\infty\sum_{m^2=1}^\infty\frac{Z_{00}(0,n^2)Z_{00}(0,m^2)}{nmL^2}\nonumber\\
&&\hspace{0cm}\times
\int_{-\infty}^\infty p^\epsilon dp\,e^{inpL}\frac{p^3F(p^2)}{E_p^2}\int_{-\infty}^\infty k^\epsilon dk\,e^{imkL}\frac{k^3}{E_k}\frac{1}{k^2-p^2}\\
&=& \frac{10}{9}\frac{\alpha^2}{36\pi^2}\sum_{n^2=1}^\infty
\frac{Z_{00}(0,n^2)}{nL}\int_{-\infty}^\infty  dp\,\sin{(npL)}\,\frac{p^3F(p^2)}{E_p^2}
\,B(p^2)\ ,
\nonumber
\end{eqnarray}
where $B(p^2)$ was defined in Eq.~(\ref{2ndintegral}), and we took the limit $d\to 3$ in the
last step.   We find that, for the parameter values of Table I, 
\begin{equation}
\label{DamuNNLOnum4}
\Delta a_\mu^{\rm HVP,\,NNLO,\,4}=\left\{\begin{array}{c} 0.02\times 10^{-10}\ ,\quad L/a=96\\
0.02\times 10^{-10}\ ,\quad L/a=64\\
0.01\times 10^{-10}\ ,\quad L/a=48
\end{array}\right.\ .
\end{equation}
The total NNLO contribution thus adds up to 
\begin{equation}
\label{DamuNNLOnumtotal}
\Delta a_\mu^{\rm HVP,\,NNLO}=\left\{\begin{array}{c} 9.13\times 10^{-10}\ ,\quad L/a=96\\
9.01\times 10^{-10}\ ,\quad L/a=64\\
7.40\times 10^{-10}\ ,\quad L/a=48
\end{array}\right.\ .
\end{equation}

\bibliographystyle{apsrev4-1}
\bibliography{ref.bib}

\end{document}